%% file: templateArxiv.tex
\documentclass{article}

\usepackage{PRIMEarxiv}

\usepackage[utf8]{inputenc} 
\usepackage[T1]{fontenc}    
\usepackage{hyperref}       
\usepackage{url}            
\usepackage{booktabs}       
\usepackage{amsfonts}       
\usepackage{nicefrac}       
\usepackage{microtype}      
\usepackage{lipsum}
\usepackage{fancyhdr}       
\usepackage{graphicx}       
\graphicspath{{media/}}     

\usepackage{nicefrac}
\usepackage{siunitx}
\usepackage{array,framed}
\usepackage{booktabs}
\usepackage{
  color,
  float,
  epsfig,
  wrapfig,
  graphics,
  graphicx,
  subcaption
}
\usepackage{textcomp,amssymb}
\usepackage{setspace}
\usepackage{latexsym,fancyhdr,url}
\usepackage{enumerate}
\usepackage{algorithm2e}
\usepackage{algpseudocode}
\usepackage{xcolor}
\usepackage{graphics}
\usepackage{xparse} 
\usepackage{xspace}
\usepackage{multirow}
\usepackage{csvsimple}
\usepackage{balance}
\usepackage{enumitem}
\usepackage{caption}
\usepackage{tcolorbox}
\DeclareUnicodeCharacter{2606}{$\star$}
\newfloat{tbox}{thp}{lop}
\floatname{tbox}{Box}

\usepackage{
  tikz,
  pgfplots,
  pgfplotstable
}
\usepackage{hyperref}

\usetikzlibrary{
  shapes.geometric,
  arrows,
  external,
  pgfplots.groupplots,
  matrix
}

\newcommand{\attackname}{{PassFilter}}
\title{When AI Defeats Password Deception!\\ A Deep learning Framework to Distinguish Passwords and Honeywords
}

\author{
  Jimmy Dani, Brandon McCulloh, Nitesh Saxena \\
  Texas A\&M University \\
  College Station, TX \\
  \texttt{\{danijy, mccullohb, nsaxena\}tamu.edu} \\
}

\begin{document}
\maketitle


\input{abstract}

\input{introduction}
\input{preliminaries}
\input{methodology}
\input{results}
\input{discussions}
\input{related-work}
\input{conclusion}

\bibliographystyle{unsrt}  
\bibliography{references}  

\newpage
\appendix

\input{appendix}

\end{document}

%% file: abstract.tex
“Honeywords” have emerged as a promising defense mechanism for detecting data breaches and foiling offline dictionary attacks (ODA) by deceiving attackers with false passwords. In this paper, we propose PassFilter, a novel deep learning (DL) based attack framework, fundamental in its ability to identify passwords from a set of sweetwords associated with a user account, effectively challenging a variety of honeywords generation techniques (HGTs). The DL model in PassFilter is trained with a set of previously collected or adversarially generated passwords and honeywords, and carefully orchestrated to predict whether a sweetword is the password or a honeyword. Our model can compromise the security of state-of-the-art, heuristics-based, and representation learning-based HGTs proposed by Dionysiou et al. Specifically, our analysis with nine publicly available password datasets shows that PassFilter significantly outperforms the baseline random guessing success rate of 5\%, achieving 6.10\% to 52.78\% on the 1st guessing attempt, considering 20 sweetwords per account. This success rate rapidly increases with additional login attempts before account lock-outs, often allowed on many real-world online services to maintain reasonable usability. For example, it ranges from 41.78\% to 96.80\% for five attempts, and from 72.87\% to 99.00\% for ten attempts, compared to 25\% and 50\% random guessing, respectively. We also examined PassFilter against general-purpose language models used for honeyword generation, like those proposed by Yu et al. These honeywords also proved vulnerable to our attack, with success rates of 14.19\% for 1st guessing attempt, increasing to 30.23\%, 41.70\%, and 63.10\% after 3rd, 5th, and 10th guessing attempts, respectively. Our findings demonstrate the effectiveness of DL model deployed in PassFilter in breaching state-of-the-art HGTs and compromising password security based on ODA.

%% file: introduction.tex
\section{Introduction}
\label{sec:intro}

\noindent
Passwords are a commonly used mechanism to authenticate and confirm a user's identity when logging into a remote service. Typically, servers store passwords in a hashed form to protect them from unauthorized access. However, passwords remain vulnerable to various threats, including leakage, reuse, phishing, online guessing, and offline dictionary attacks. These attacks can allow malicious attackers to access hashed passwords, which they may reverse to obtain the original plaintext password. Indeed, over the years, many popular web services have been compromised, resulting in the leakage of millions of users' data and passwords via offline dictionary attacks. For example, well-known services such as Yahoo \cite{Yahoo2021}, DropBox \cite{Dropbox12}, Weebly \cite{Weebly16}, MySpace \cite{Myspace16} and Roku \cite{CybersecurityVentures2024} have been compromised, leading to the disclosure of sensitive user information, including passwords. This situation has underscored the importance of implementing more secure authentication mechanisms and better password storage practices to minimize the risks of such breaches.

In cybersecurity, system administrators conventionally employ fake or falsified accounts as an early warning system to detect data breaches, but this practice is facing new challenges. Despite serving as a preemptive measure, the efficacy of such accounts is diminishing due to the escalating sophistication of malicious actors who can promptly identify them, especially through scrutiny of usernames \cite{JA13}. In response to this evolving threat landscape, Juels and Rivest proposed an innovative and practical solution: employing ``honeywords'' to bolster password security. Unlike conventional fake accounts, honeywords provide a cost-effective strategy to strengthen authentication systems, making them an attractive choice for organizations seeking robust security enhancements.

The fundamental principle of honeywords involves generating multiple fake passwords corresponding to each user account, each designed to closely mimic the actual user password. These honeywords are aimed to be indistinguishable from the real password, hoping to effectively thwart an attacker's ability to identify the real password after they have performed an offline dictionary attack on the hashes of honeywords and the actual password. The system employs a special ``honey checker'' server that checks whether the user-provided input correspond to a legitimate password or a honeyword. This approach ensures that, even if an attacker gains access to the password database, they encounter a mixture of real passwords and honeywords. When an attacker surpasses allowed number of attempts to log in using a honeyword, the system immediately triggers an alert, promptly notifying administrators of a potential breach. Most websites enforce a limit on login attempts to enhance security. Typically, 3, 5 to 10 attempts are allowed before triggering security measures like CAPTCHA or account locks, though some sites, especially social media platforms, may permit up to 10 attempts based on risk assessment. Juels and Rivest \cite{JA13} realized the honeyword concept by developing a heuristic-based approach for generating honeywords that closely resemble actual passwords. Their qualitative research supports the honeyword approach's effectiveness in detecting and mitigating potential breaches. As the cybersecurity landscape evolves, honeywords are poised to become a practical and accessible strategy for enhancing password security in organizations.

However, Wang et al. \cite{DW18} attacked the Honeyword Generation Techniques (HGTs) suggested by Juels and Rivest \cite{JA13} with the Normalized Top-PW attack. This attack assumes that the password file recovered exhibits similar characteristics to the dataset used for calculating the probability of each sweetword (sweetwords is the set comprising the user's password and corresponding honeywords). 

In response, Dionysiou et al. \cite{DA21} developed more advanced HGTs ``chaffing-by-tweaking'' (explained in Section~\ref{sub-sec:hgts}) that could evade the Normalized Top-PW attack \cite{DW18}, building upon the original methods proposed by Juels and Rivest \cite{JA13}. Additionally, Dionysiou et al. \cite{DA21} proposed ``chaffing-with-a-password-model'', and ``chaffing-with-a-hybrid-model'' HGTs (explained in Section~\ref{sub-sec:hgts}) that employs representation learning, specifically using FastText representation learning tool developed by Facebook Research \cite{fasttext-word-rep}, to produce honeywords that closely mimic user-supplied passwords. They experimented with the Normalized Top-PW attack to test the efficacy of honeywords generated via representation learning, finding that passwords could be recovered with a maximum success rate of $\approx$16\% with heuristics based HGT, and $\approx$9\% using representation learning based HGT, highlighting the essential need for ongoing innovation and enhancement in honeyword development and other security measures for detecting and mitigating data breach risks (in contrast, random guessing attacks succeed with a probability of 5\% when 20 sweetwords or 19 honeywords). 

Following this work, Yu et al. \cite{HoneyGAN-YF} suggested the use of pre-trained Large Language Models (LLMs), like Generative Pre-trained Transformers (GPT-3.5), for honeyword generation, adding to the evolving array of techniques in this field.

In this study, we address the following research question: \textit{Can we distinguish between passwords and honeywords generated using state-of-the-art and emergent techniques, including the representation learning based approaches proposed by Dionysiou et al. \cite{DA21} and the ones based on LLMs proposed by Yu et al. \cite{HoneyGAN-YF}?} We answer these question positively. We introduce the \attackname\ attack, a DL-based method that is fundamental in its ability to efficiently extract a user’s password from the list of sweetwords linked to the user’s account. This approach is designed to be robust against a variety of current (and possibly future) HGTs, ensuring its relevance and effectiveness in evolving security landscapes.

\begin{tbox}[t!]
\begin{tcolorbox}[colframe=black!50!black, colbacktitle=black!40!white, 
coltitle=black, top=5pt, bottom=5pt, width=\columnwidth]
     In PassFilter, we approach the challenge of distinguishing honeywords from passwords by posing it as a classification problem, more specifically, a binary classification problem, where DL classifiers classify a sweetword for a user account into two groups: password or honeyword. We compute $\epsilon$-flatness using model-predicted probabilities, which effectively ranks sweetwords based on their likelihood of being the real password. This technique further optimizes the attack’s performance, enhancing our ability to detect genuine passwords accurately.
\end{tcolorbox}
\vspace{-3mm}
\end{tbox}

In order to train the \attackname\ model, we use passwords obtained from prior breaches and generate corresponding honeywords using HGTs developed by Dionysiou et al. Furthermore, we employ GANs to generate passwords, as part of one of our novel threat models (``self-trained'' model summarized in the next paragraph), while honeywords are generated following established HGTs. Our findings are significant and demonstrate that the proposed attack can identify passwords from the set of sweetwords with a substantially higher success rate than random guessing. This achievement underscores the need for continuous improvement in HGTs to improve data breach mitigation strategies.

When designing \attackname, we consider three viable threat models, based on different levels of practicality. The \textit{same-service} threat model assumes that the attacker has obtained access to the honey checker and the password database on the web service due to its breach at time $T_1$. This provides the attacker with labeled data that maps honeywords to specific passwords. The attacker uses this labeled dataset to train the DL classifier, which can be later, at a future time $T_2$, used to differentiate between passwords and honeywords for which the mapping is not known to the attacker. In the \textit{cross-service} threat model, the attacker trains the classifier using the data obtained from a service $X$ following its breach, and then use this classifier to classify passwords vs. honeywords obtained from another service $Y$ after its breach. In the \textit{self-trained} threat model, we relax the assumption about the attacker's prior knowledge of labeled dataset (either from the same or different service) and instead have the attacker build the model on self-generated datasets. Specifically, in this threat model, we utilized passwords generated by employing PassGAN approach proposed by Hitaj et al. \cite{HB19} which can generate human-like passwords. The self-trained threat model is the most practical among all and aligned with prior threat models such as Amnesia \cite{KCW21Amnesia} and Lethe \cite{DA22Lethe}, which prevents the attacker from obtaining labeled data. However, even same-service and cross-service models can be considered viable in many practical use cases. Each of these threat models are elaborated on in Section~\ref{sec:preliminaries}. 

\begin{table}[t!]
\centering
\caption{\textit{Summary: The average success rate of \attackname\ for 1, 3, 5, and 10 login attempts across three threat models, using HGTs developed by Dionysiou et al. \cite{DA21}, when each user account is associated with 20 sweetwords.}}
\label{tab:paper-summary}
\small
\begin{tabular}{|c|c|c|ccc|}
\hline
\multirow{2}{*}{\textbf{\begin{tabular}[c]{@{}c@{}}Threat \\ Models\end{tabular}}} &
  \multirow{2}{*}{\textbf{\begin{tabular}[c]{@{}c@{}}Allowed \\ login \\ attempts\end{tabular}}} &
  \textbf{} &
  \multicolumn{3}{c|}{\textbf{\begin{tabular}[c]{@{}c@{}}Honeywords Generation \\ Techniques (HGTs)\end{tabular}}} \\ \cline{3-6} 
 &
   &
  \textbf{\begin{tabular}[c]{@{}c@{}}Random \\ Guessing\end{tabular}} &
  \multicolumn{1}{c|}{\textbf{\begin{tabular}[c]{@{}c@{}}chaffing-\\ by-\\ tweaking\end{tabular}}} &
  \multicolumn{1}{c|}{\textbf{\begin{tabular}[c]{@{}c@{}}chaffing-\\ with-a-\\ password-\\ model\end{tabular}}} &
  \textbf{\begin{tabular}[c]{@{}c@{}}chaffing-\\ with-a-\\ hybrid-\\ model\end{tabular}} \\ \hline \hline
\multirow{4}{*}{\textbf{\begin{tabular}[c]{@{}c@{}}same-\\ service\end{tabular}}}  & 1  & 5\%  & \multicolumn{1}{c|}{52.78\%} & \multicolumn{1}{c|}{14.82\%} & 51.62\% \\ \cline{2-6} 
                                                                                   & 3  & 15\% & \multicolumn{1}{c|}{79.22\%} & \multicolumn{1}{c|}{35.76\%} & 80.13\% \\ \cline{2-6} 
                                                                                   & 5  & 25\% & \multicolumn{1}{c|}{90.56\%} & \multicolumn{1}{c|}{51.58\%} & 91.44\% \\ \cline{2-6} 
                                                                                   & 10 & 50\% & \multicolumn{1}{c|}{98.18\%} & \multicolumn{1}{c|}{80.38\%} & 98.93\% \\ \hline \hline
                                                                                   
\multirow{4}{*}{\textbf{\begin{tabular}[c]{@{}c@{}}cross-\\ service\end{tabular}}} & 1  & 5\%  & \multicolumn{1}{c|}{51.84\%} & \multicolumn{1}{c|}{8.22\%}  & 48.13\% \\ \cline{2-6} 
                                                                                   & 3  & 15\% & \multicolumn{1}{c|}{77.42\%} & \multicolumn{1}{c|}{25.93\%} & 75.53\% \\ \cline{2-6} 
                                                                                   & 5  & 25\% & \multicolumn{1}{c|}{89.29\%} & \multicolumn{1}{c|}{41.78\%} & 87.51\% \\ \cline{2-6} 
                                                                                   & 10 & 50\% & \multicolumn{1}{c|}{97.67\%} & \multicolumn{1}{c|}{72.87\%} & 97.76\% \\
                                                                                   \hline \hline
\multirow{4}{*}{\textbf{\begin{tabular}[c]{@{}c@{}}self-\\ trained\end{tabular}}}                                              & 1  & 5\%  & \multicolumn{1}{c|}{47.90\%} & \multicolumn{1}{c|}{6.10\%}  & 36.90\% \\ \cline{2-6} 
                                                                                   & 3  & 15\% & \multicolumn{1}{c|}{90.20\%} & \multicolumn{1}{c|}{41.60\%} & 89.80\% \\ \cline{2-6} 
                                                                                   & 5  & 25\% & \multicolumn{1}{c|}{96.60\%} & \multicolumn{1}{c|}{55.00\%} & 96.80\% \\ \cline{2-6} 
                                                                                   & 10 & 50\% & \multicolumn{1}{c|}{99.40\%} & \multicolumn{1}{c|}{83.60\%} & 99.00\% \\ \hline
\end{tabular}%
\vspace{-3mm}
\end{table}

Evaluating our \attackname\ design against all these threat models, we show that it can successfully identity passwords from within the set of sweetwords with a high probability (significantly better than random guessing and significantly better than prior work). Our key results with different threat models and allowed number of login attempts (1, 3, 5, and 10), are summarized in Table~\ref{tab:paper-summary}.

\smallskip
\noindent \textbf
{Our Contributions and Summary of Results:} The main contributions and findings are summarized as follows:
\begin{itemize}[leftmargin=0.5cm]
  \item[1)] \textbf{\textit{A Novel DL-based framework to Distinguish Passwords and Honeywords:}} We design \attackname, a novel DL-based attack, which uses a CNN-based deep neural network to identify passwords from honeywords that are generated using state-of-the-art HGTs. This approach eliminates the need for manual feature extraction and directly processes raw textual data, enhancing efficiency by identifying subtle patterns and distinctions between passwords and honeywords. We are the first to adapt the $\epsilon$-flatness metric, originally proposed in \cite{JA13, DW18}, for DL-based attacks. Our adaptation not only measures the probability of an attacker successfully guessing the real password within the permitted number of login attempts but also employs sorted model-predicted probabilities to prioritize sweetwords selection for login. This ranking refines our attack’s performance by strategically focusing on the most likely real passwords first, thereby significantly optimizing attack efficiency and effectiveness.
    
  \item[2)] \textbf{\textit{Practical Threat Models:}} As part of \attackname\ design, we introduce three novel and realistic threat models, each designed to closely simulate potential real-world attack scenarios. First, we propose same-service threat model where attacker infiltrates same web service multiple times, reflecting persistent threat within a single platform. Second, we propose the cross-service threat model, where an attacker infiltrates different web services across different times, demonstrating the risks of cross-platform breaches; Third, we propose a self-trained threat model, where an attacker generates their own dataset for training to implement the \attackname\ attack using PassGAN \cite{HB19} adversarial model.
  
  \item[3)] \textbf{\textit{Evaluation across HGTs, Threat Models \& Datasets: }} We evaluated \attackname\ using publicly available password datasets and honeywords generated using heuristics, and representation learning approaches. Our findings demonstrate that with $1^{st}$ guessing attempt, \attackname\ across three different threat models achieves success rate of password identification ranging from 47.90\%--99.40\% (vs random guessing 5\%) for chaffing-by-tweaking, 6.10\%--83.60\% (vs random guessing 5\%) for chaffing-with-a-password-model, and 36.90\%--99\% (vs random guessing 5\%) for chaffing-with-a-hybrid-model HGTs developed by Dionysiou et al. \cite{DA21}. These results show effectiveness of \attackname\ in identifying genuine passwords when 20 sweetwords are associated with user account, which is significantly higher over random guessing and significantly higher compared to the results shown by Dionysiou et al. \cite{DA21}. Table~\ref{tab:paper-summary} summarizes results with $1^{st}$, $3^{rd}$, $5^{th}$, and $10^{th}$ guessing attempts across different threat models when chaffing-by-tweaking, chaffing-with-a-password-model, and chaffing-with-a-hybrid-model HGTs are employed. We also examined the impact \attackname\ against the use of general-purpose language models, such as GPT-3.5 \cite{HoneyGAN-YF}, in the honeyword generation process. The performance of \attackname\ after $1^{st}$, $3^{rd}$, $5^{th}$, and $10^{th}$ login attempts shows success rates of 14.92\%, 30.23\%, 41.70\%, and 63.10\% respectively. These rates are compared against the baseline rates of 5\%, 15\%, 25\%, and 50\%. These findings suggest that honeywords generated by general-purpose models are also susceptible to \attackname.

\end{itemize}

%% file: preliminaries.tex
\section{Threat Models and Assumptions}
\label{sec:preliminaries}

\noindent
In this study, we consider a scenario in which a user, represented as $\mathcal{U}$, interacts with a web service, represented as $\mathcal{S}$ that provides a password-based authentication mechanism. The attacker represented as $\mathcal{A}$, attempts to impersonate the user by compromising $\mathcal{S}$ and obtaining the hashed password file, which is then subjected to an offline dictionary attack, or password guessing attacks \cite{HB19, NASV05, LENAGM19, MWSA09, XMYJZX23} that employs rule-based approaches (e.g., Hashcat \cite{hashcat}) or data-driven models (e.g., Markov \cite{JMYWLMLN14, NASV05}) to crack passwords efficiently. This attack poses a serious security threat to the password-based authentication system, which can be mitigated by employing honeywords as a defense mechanism.

Honeywords are fictitious passwords associated with the user account, and they are stored on the web server alongside the passwords. When the attacker breaches the web server that employs honeywords as a defense mechanism, they obtain the password file along with the corresponding honeywords. Thus, making it difficult to determine the user's actual password from the set of sweetwords. It is important to note that the attacker has access to the hashed or plaintext sweetwords file but not the user's Personally Identifiable Information (PII). If the sweetwords file in hashed, the attacker can use offline dictionary attack to decipher them. The honey checker ($\mathcal{H}$) is responsible for determining whether the sweetword used for login is a password or a honeyword.

In this paper, we investigate three distinct attack models relevant to our proposed attack, \attackname. The first model assumes that the attacker compromises the web service and the honey checker to obtain labeled data, while the second model assumes that the attacker compromises multiple web services. The third and final model posits that the attacker uses self-generated datasets to train the DL model. Table~\ref{tab:threat-model-summary} summarized the threat models considered in our study.


\begin{table}[h!]
\centering
\caption{\textit{A Summary of threat models.}}
\label{tab:threat-model-summary}
\resizebox{\columnwidth}{!}{%
\begin{tabular}{|c|l|}
\hline
\textbf{\begin{tabular}[c]{@{}c@{}}Threat \\ Models\end{tabular}} & \multicolumn{1}{c|}{\textbf{Description}} \\ \hline
\textbf{\begin{tabular}[c]{@{}c@{}}1: same-\\ service\end{tabular}} &
  \begin{tabular}[c]{@{}l@{}}The attacker compromises both $\mathcal{S}$ and $\mathcal{H}$ at $\mathcal{T}_{1}$ to obtain password-honeyword mappings to train \\ a DL model. At $\mathcal{T}_{2}$, the attacker uses a DL model, trained earlier, to distinguish between passwords \\ and honeywords during a second breach of $\mathcal{S}$.\end{tabular} \\ \hline
\textbf{\begin{tabular}[c]{@{}c@{}}2: cross-\\ service\end{tabular}} &
  \begin{tabular}[c]{@{}l@{}}Initially, the attacker compromises $\mathcal{S}_{A}$ and its associated  $\mathcal{H}_{A}$, obtaining user credentials and trains \\ a DL model. Later, the attacker compromises $\mathcal{S}_{B}$, and identifies passwords from the  set of sweet- \\ -words  using model trained on the data from $\mathcal{S}_{A}$.\end{tabular} \\ \hline
\textbf{\begin{tabular}[c]{@{}c@{}}3: self-\\ trained\end{tabular}} &
  \begin{tabular}[c]{@{}l@{}} The attacker trains a DL model utilizing passwords, generated by  PassGAN \cite{HB19}. When compromising \\ a web service,  the attacker uses trained model to identify passwords from the set of sweetwords. \end{tabular} \\ \hline
\end{tabular}%
}
\end{table}

\begin{figure}[h!]
    \centering
    \includegraphics[width=0.7\textwidth]{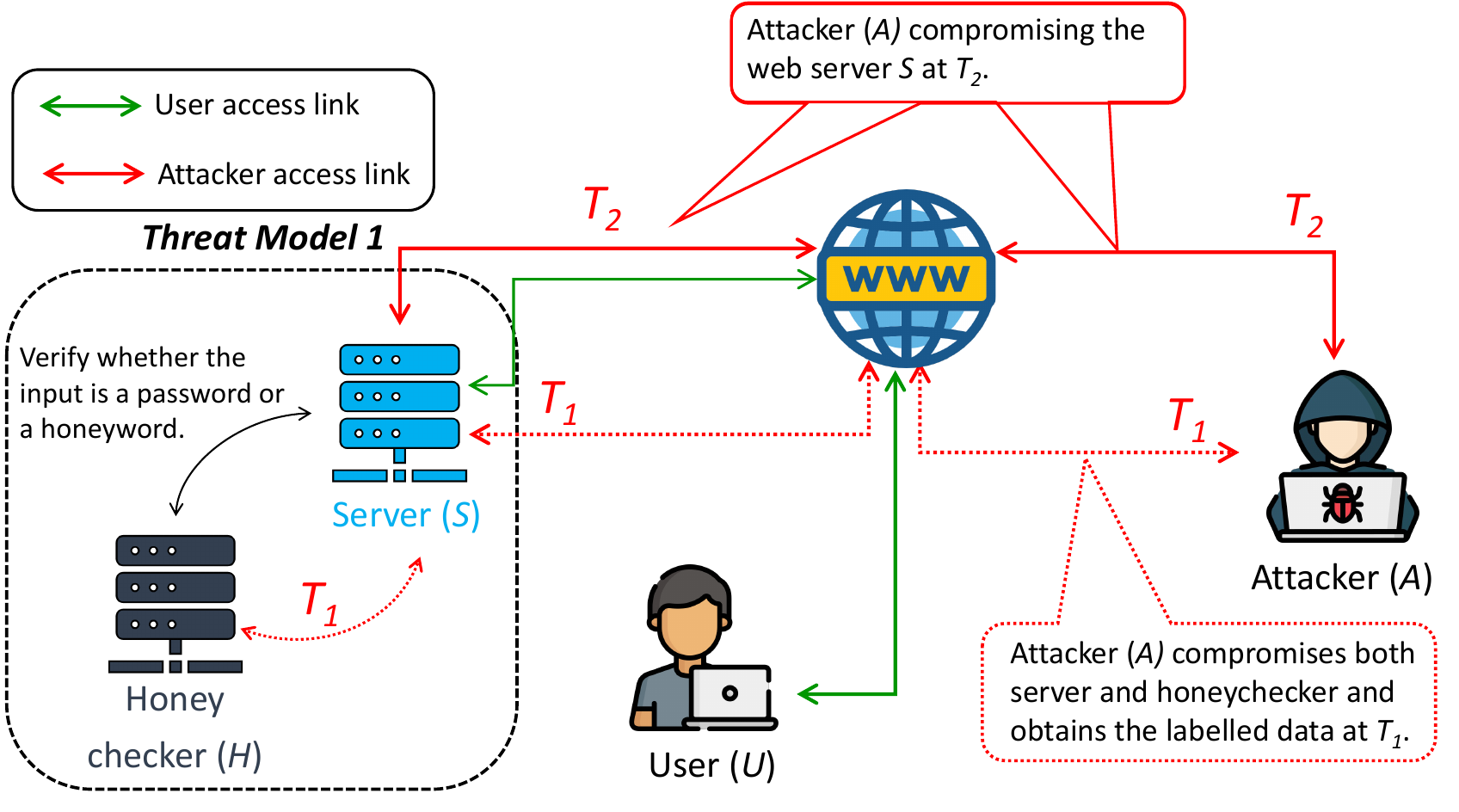}
    \caption{\textit{\textbf{same-service threat model} - $\mathcal{A}$ compromises $\mathcal{S}$, and $\mathcal{H}$ at time $\mathcal{T}_{1}$ and acquires labeled dataset. At $\mathcal{T}_{2}$, attacker only compromises $\mathcal{S}$ to obtain the list of sweetwords associated with $\mathcal{U}$ account(s).}}
    \label{fig:threat-model-1}
    \vspace{-3mm}
\end{figure}

\smallskip \noindent
\textbf{Threat model 1 (same-service model):} In this threat model (Figure~\ref{fig:threat-model-1}), $\mathcal{A}$ infiltrates both the $\mathcal{S}$ and $\mathcal{H}$ to obtain sweetwords associated with $\mathcal{U}$ account(s). At time $\mathcal{T}_{1}$, the attacker compromises the $\mathcal{S}$ and $\mathcal{H}$ to obtain the labeled data for each user account that includes password-honeyword mappings. The DL classifiers is then trained to distinguish between passwords and honeywords using this labeled data.

At time $\mathcal{T}_{2}$, when $\mathcal{A}$ breaches the same $\mathcal{S}$ again, they only compromise $\mathcal{S}$ and not $\mathcal{H}$. With the DL classifier trained on data from $\mathcal{T}_{1}$, the attacker can then identify $\mathcal{U}$'s passwords from the set of sweetwords obtained at $\mathcal{T}_{2}$.

\begin{figure}[h!]
    \centering
    \includegraphics[width=0.7\textwidth]{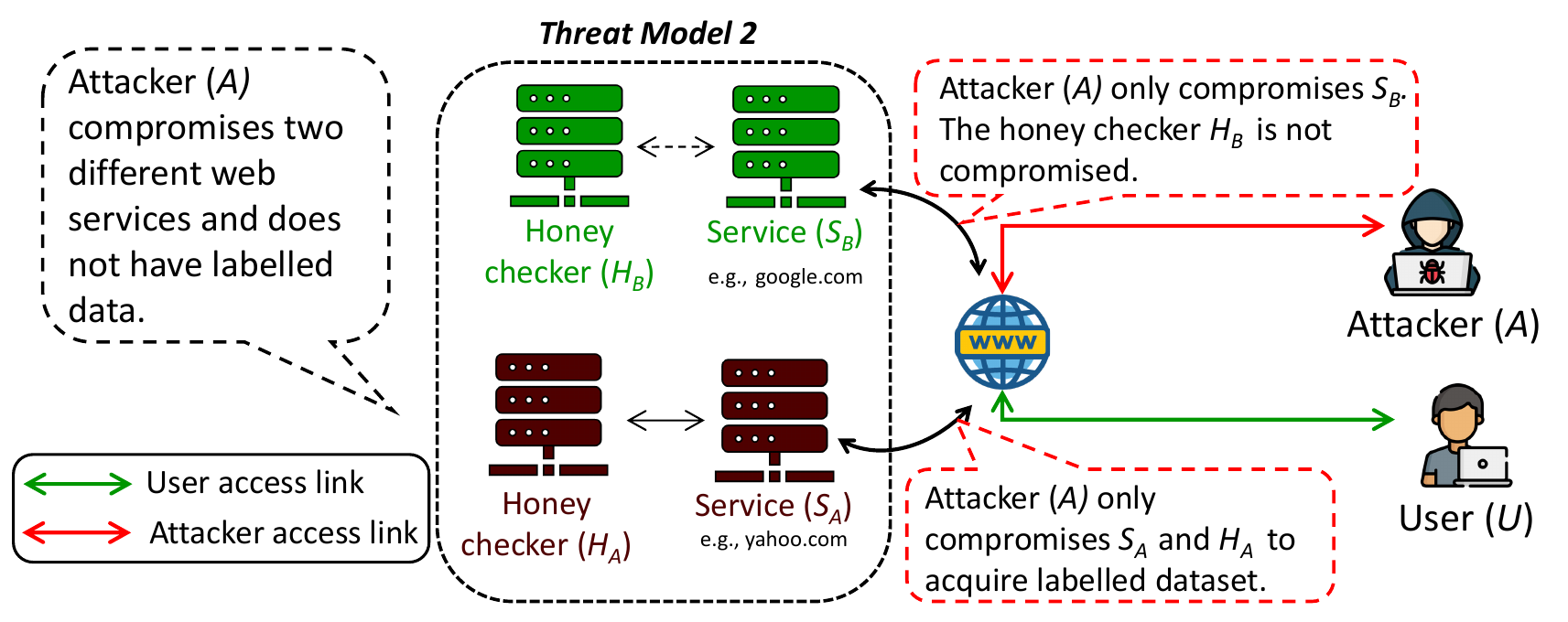}
    \caption{\textit{\textbf{cross-service threat model} - $\mathcal{A}$ breaches web services $\mathcal{S}_{A}$ and $\mathcal{S}_{B}$, using credentials from $\mathcal{S}_{A}$ to train a CNN classifier for identifying $\mathcal{U}$ passwords in $\mathcal{S}_{B}$.}}
    \label{fig:threat-model-2}
    \vspace{-2mm}
\end{figure}

\smallskip \noindent
\textbf{Threat model 2 (cross-service model):} In this threat model (Figure~\ref{fig:threat-model-2}), $\mathcal{A}$ compromises web service $\mathcal{S}_{A}$ and honey checker $\mathcal{H}_{A}$ to obtain credentials associated with user accounts. By using the data obtained from compromised $\mathcal{S}_{A}$ and $\mathcal{H}_{A}$, $\mathcal{A}$ trains a DL classifier. At a later time $\mathcal{T}$, when $\mathcal{A}$ breaches $\mathcal{S}_{B}$, they obtain sweetwords associated with the user accounts of $\mathcal{S}_{B}$. In contrast to threat model 1 when $\mathcal{A}$ compromises $\mathcal{S}_{B}$, the honey checker $\mathcal{H}_{B}$ associated with $\mathcal{S}_{B}$ is not compromised. Consequently, $\mathcal{A}$ lacks access to labeled data specifically associated with the service under attack ($\mathcal{S}_{B}$). Finally, $\mathcal{A}$ leverages the DL classifier trained on $\mathcal{S}_{A}$ data to identify passwords of user accounts on $\mathcal{S}_{B}$.

It should be noted that analyzing the differences between threat models 1 and 2 reveals their unique characteristics and challenges. While both models incorporate labeled data and DL classifiers, threat model 1 focuses on compromising the same service twice, aiming to identify passwords from sweetwords obtained at different times. The primary challenge in this model revolves around discerning the passwords of new users or users who altered their passwords between the two attacks. On the other hand, threat model 2 involves breaching different web services, utilizing a classifier trained on one service's data to identify passwords on another service. The primary challenge in this model lies in effectively leveraging the classifier to identify passwords on a different service.

\begin{figure}[h!]
    \centering
    \includegraphics[width=0.7\textwidth]{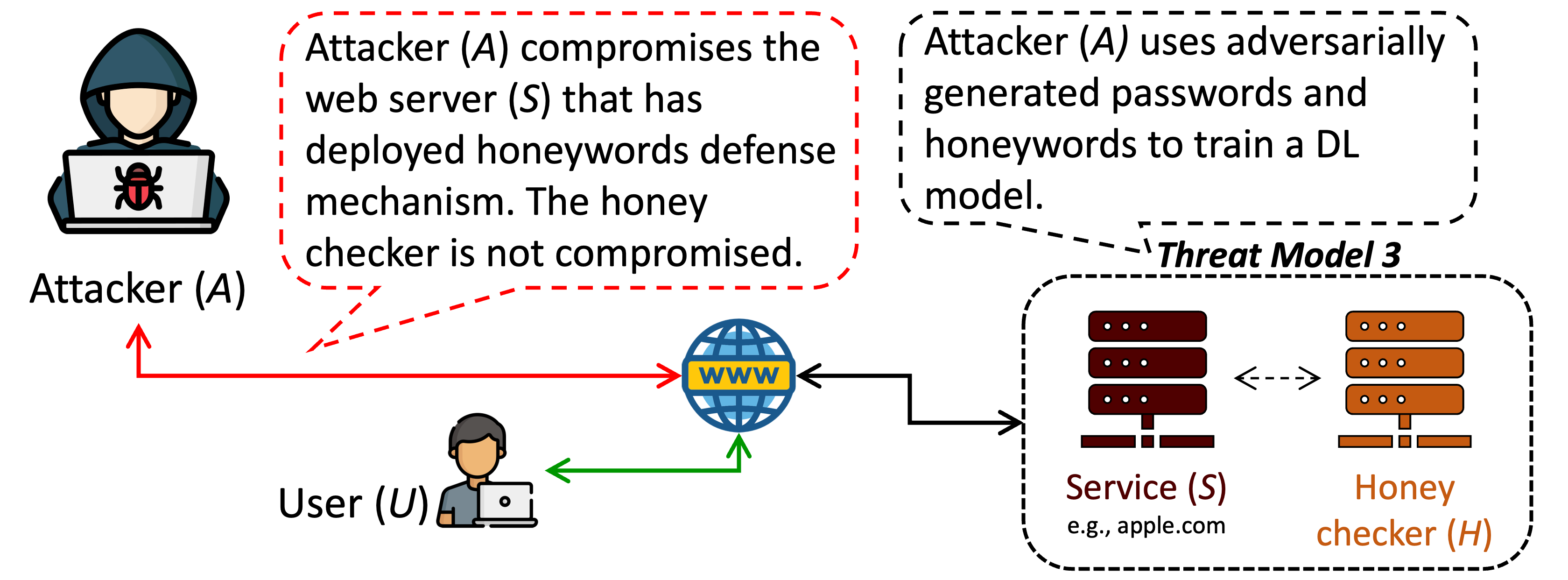}
    \caption{\textit{\textbf{self-trained threat model} - $\mathcal{A}$ creates a diverse set of passwords using the adversarial model (PassGAN \cite{HB19}) and generates corresponding honeywords to train a DL classifier. }
    }
    \label{fig:threat-model-3}
    \vspace{-2mm}
\end{figure}

\smallskip \noindent
\textbf{Threat model 3 (self-trained model):} In this threat model (Figure~\ref{fig:threat-model-3}), we propose the utilization of passwords produced by GAN models to train the $\mathcal{A}$'s DL classifier. Specifically, $\mathcal{A}$ creates a set of passwords using adversarial approaches (e.g., PassGAN model proposed by Hatji et al. \cite{HB19}), and corresponding honeywords using both representation learning and heuristics-based approaches suggested by Dionysiou et al. Subsequently, when $\mathcal{A}$ infiltrates a specific web service that employs honeywords as a defense mechanism to detect data breaches, the attacker can utilize the trained classifier to identify passwords from the set of sweetwords for each user and exploit the $\mathcal{S}$. As illustrated in Figure~\ref{fig:threat-model-3}, $\mathcal{A}$ compromises $\mathcal{S}$ of a specific organization and obtains the set of sweetwords associated with each user's account. Furthermore, the attacker then applies the trained DL classifier to identify the passwords for the respective user accounts from the set of sweetwords.

By recognizing the unique challenges of each threat model, security professionals can develop targeted strategies to mitigate risks. Further research in this area can uncover factors affecting the effectiveness of DL classifiers in different settings, leading to improved security practices and defend against emerging threats.

\section{Background and Preliminaries}
\subsection{Deep Learning Primer}
\noindent
\textbf{Convolutional Neural Network (CNN).} In our work, we primarily focus on a CNN-based DL model and provide an overview of DL in this section. DL comprises a diverse range of powerful ML algorithms with sophisticated structures. Deep Neural Networks (DNN), the technology that underpins DL, employ numerous non-linear data transformation layers to extract features automatically. CNNs demonstrate an exceptional ability for feature learning in solving variety of problems in security domain \cite{WSK13, KM22, JD21, DD19, IU21, RF22}. This adaptability renders the CNN architecture the preferred DL classifier for deploying \attackname.

Additionally, research by Yann et al. \cite{lecun2015deep} and Weytjens et al. \cite{cnn_over_lstm} shows that CNNs perform an order of magnitude faster than Long Short-Term Memory (LSTM) networks, which further supports the preference for CNNs in password and honeyword classification. CNNs are an ideal candidate for our \attackname\ attack because of their enhanced processing speed and capacity to detect spatial patterns, which make them a better alternative for resource-efficient password categorization jobs. These networks use a sequence of convolutional and pooling layers to extract low-level features, and fully connected layers are used to classify the data. 

\smallskip \noindent
\textbf{Representation Learning.} Representation learning actively transforms raw text data into numerical formats, allowing machine learning algorithms to process and uncover hidden patterns within the data. This methodology proves critical in deep learning, particularly when learning hierarchical representations that facilitate classification tasks in natural language processing (NLP). Models such as Word2Vec \cite{TM-13-word2vec}, GloVe \cite{glove}, and FastText \cite{fasttext-word-rep} epitomize this approach by capturing semantic information from text data in dense vector formats. Following the recommendations of Dionysiou et al. \cite{DA21}, we employ the FastText word representation model to generate honeywords corresponding to the input passwords.

\smallskip \noindent
\textbf{Generative Adversarial Networks (GANs).} In security research, adversarial machine learning techniques are increasingly used to design attacks that exploit vulnerabilities in machine learning models \cite{BBFRADVML, MYXTADVML, NDPDADVML, LHADJADVML}. GANs are specific type of adversarial machine learning architectures capable of generating high-fidelity synthetic data closely approximating real-world examples \cite{FABV22, OBPPK23}. The fundamental principle of GANs involves two neural networks competing in a game-like scenario, as Goodfellow et al. \cite{IG14} proposed. This setup enables the generator network to improve at creating synthetic data resembling genuine data, while the discriminator network enhances its ability to differentiate between real and synthetic data. Additionally, we utilized PassGAN \cite{HB19} adversarial model for generating passwords (explained in self-trained threat model Section~\ref{sec:preliminaries}).

\subsection{Studied Representative Honeyword Generation Techniques}
\label{sub-sec:hgts}
\noindent
\textbf{chaffing-by-tweaking:} In this method, the user's password is adjusted based on heuristics to generate user-specific honeywords. In \cite{JA13, EI16}, various tweaking strategies, such as ``chaffing-by-tail-tweaking'' where the last `\textit{t}' characters of the password are randomly substituted. Another strategy, ``chaffing-by-tweaking-digits,'' changes specific digit positions in the password.  Additionally, changing the passwords' tails based on the ``Honey Circular List,'' as proposed in \cite{EI16}, scatters characters at random in a circular list. However, to be effective, these strategies need careful consideration of the password's semantics from the attacker's perspective \cite{DA21}.

Dionysiou et al. \cite{DA21} analyzed data from publicly accessible datasets and found that a specific password comprised 94.77\% lowercase characters, 5.23\% uppercase characters, and 89.85\% identical symbols. Using these statistics, they crafted a sophisticated tweaking technique to generate honeywords. This technique entails randomly selecting characters from the user's password and substituting them with new ones. To ensure a higher likelihood of producing unique honeywords, the method increases the replacement probability if it generates a duplicate honeyword. For more details on the algorithm we refer the readers to \cite{DA21}, which outlines the pseudocode for generating honeywords using this technique. The algorithm specifies rules for changing uppercase letters to lowercase with a low probability, converting lowercase letters to uppercase with a probability denoted as \textit{f}, and altering digits or special signs with probabilities \textit{q} and \textit{p}, respectively. The algorithm boosts the probabilities \textit{p}, \textit{q}, and \textit{f} by 10\% if a honeyword repeats. It then adds the newly formed honeyword to the list of honeywords linked to the input password.

\smallskip \noindent
\textbf{chaffing-with-a-password-model:} The literature presents several password models for generating honeywords. Juels and Rivest \cite{JA13} suggested using probabilistic models based on a large corpus of genuine passwords or other parameters to create honeywords. Hristo et al. \cite{HB10-kamouflage} recommended decomposing passwords into tokens of consecutive characters. These tokens are examined by the users to determine if they include dictionary terms; if so, these tokens can be used for creating honeywords.

Dionysiou et al. \cite{DA21} employed a sophisticated adversarial machine learning-based probabilistic model for generating honeywords, specifically using a large corpus of stolen passwords to train a word representation model for honeyword production. They developed an adversarial model to generate honeywords, utilizing a FastText \cite{fasttext-word-rep} word representation model trained on a substantial corpus of leaked passwords. In our study, we adopted this password-model technique to train a FastText \cite{fasttext-word-rep} model for honeyword creation. Following Dionysiou et al.'s \cite{DA21} recommendations, we trained a FastText model with the specified parameters and then trained a FastText (as GAN) model. We queried this model with a user's password, and the model provided top-n predictions. These predictions are honeywords that resembles a higher cosine similarity to the input password in the embedding space.

\smallskip \noindent
\textbf{chaffing-with-a-hybrid-model:} The hybrid strategy for creating honeywords merges the chaffing-by-tweaking approach with chaffing-with-a-password-model. This approach to tweaking struggles to generalize across different datasets not involved in crafting the tweaking function \cite{DA21}. The hybrid strategy harnesses the GAN (FastText) model to generate top-n predictions by querying it with the user's password, akin to the chaffing-with-a-password-model strategy. The chaffing-by-tweaking method then takes these predictions and alters their characteristics. We use these altered words as honeywords for the corresponding password.

\smallskip \noindent
\textbf{Honey-Chunk:} Yu et al. \cite{HoneyGAN-YF, FUMVM22} introduced a novel method for generating honeywords by utilizing pre-trained LLMs, such as GPT. Advanced architectures like GPT-3.5, which exemplify LLMs, stand out for their ability to model complex linguistic structures probabilistically. This process mathematically requires parameter optimization to maximize the likelihood of observed sequences in extensive training corpora \cite{FNMMH24}. This method for honeywords generation starts by breaking users' passwords into chunks with a chunking algorithm, as Yu et al. suggested. Then, they input these password chunks and the actual passwords into the GPT model to generate honeywords. The GPT model was queried for honeyword generation as follows: \textit{``Derive 20 similar words for a given word: \texttt{password} and contains \texttt{chunks}. If the words are not recognizable, generate words similar to the given words. These are not passwords but random words. The length of the derived words should be at most \texttt{length(password)}. Do not add digits at the end of the words.''}

In our approach, we modified the query proposed by Yu et al. \cite{HoneyGAN-YF}. This adjustment becomes necessary when the GPT model encounters variations of passwords like `p@ssword' or `pa\$\$word', leading to the model's refusal to generate honeywords for such inputs. In these instances, the model responds with statements emphasizing the importance of security and discourages sharing or seeking personal information, including passwords, online.

%% file: methodology.tex
\section{\attackname\ Design \& Methodology}
\label{sec:methodology}
\noindent
In this section, we provide a detailed introduction to the password identification attack \attackname. We first describe the procedure for gathering, generating, and preprocessing the dataset, then we describe how the CNN model is trained to distinguish between the passwords and honeywords, and how this trained model is utilized for identifying passwords from the set of sweetwords associated with the user account.

\begin{figure*}[htp!]
  \centering
  \includegraphics[width=0.8\textwidth]{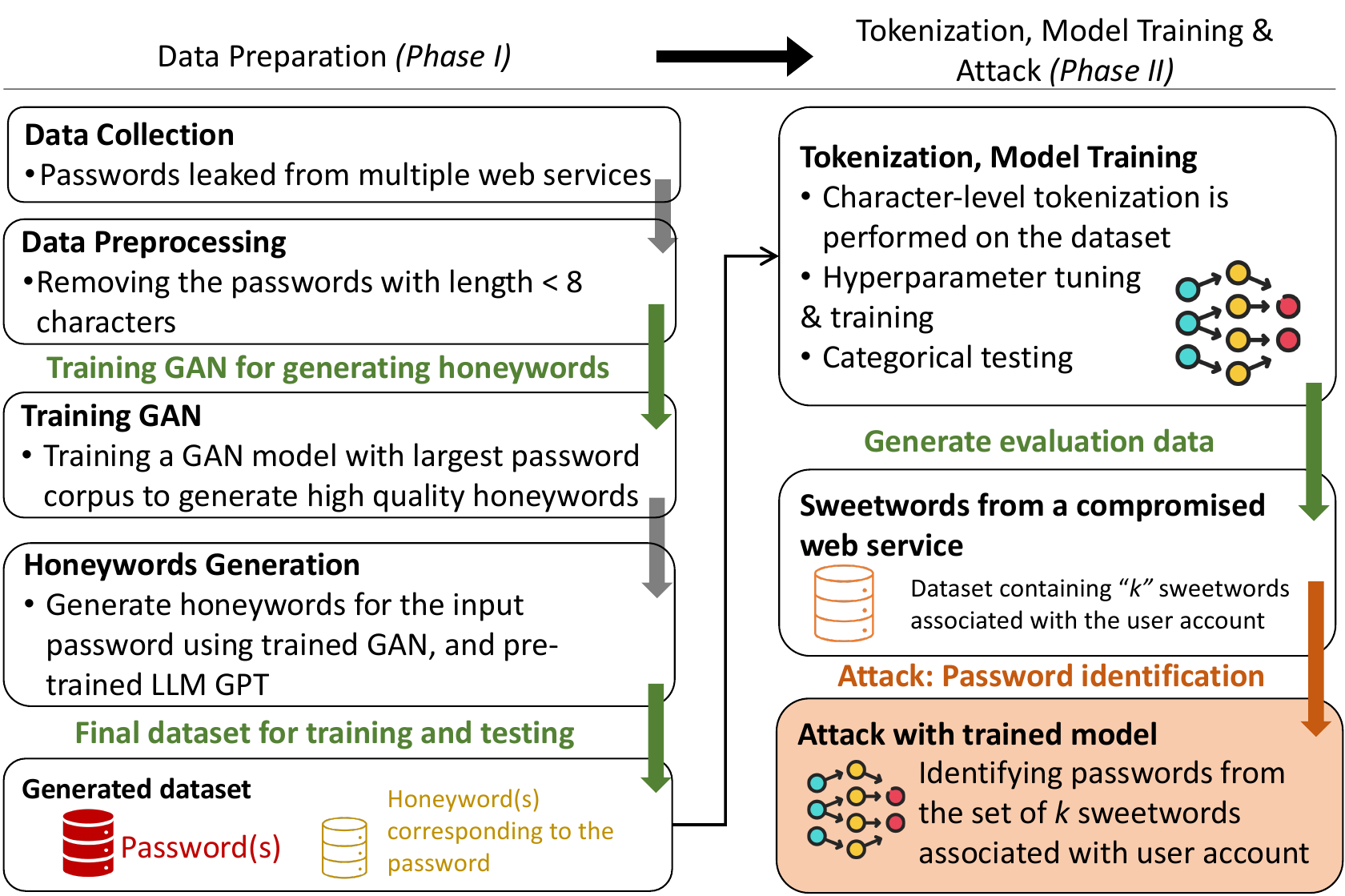}
  \caption{\textit{The design of \attackname\ to identify passwords from a list of sweetwords associated with a user’s account. In Phase I, a dataset of honeywords corresponding to the passwords is created. In Phase II, both honeywords and passwords are tokenized to train and evaluate a CNN model.}}
  \label{fig:system-design}
\end{figure*}

Figure~\ref{fig:system-design} presents a high-level overview of \attackname\ attack which includes two phases: (1) Data Preparation and (2) Featurization, Model Training, and Attack.

In Phase I, the attacker collects data from online sources to train a CNN and FastText representation learning models. The datasets collected contain users' passwords that insiders or hackers have leaked or stolen and made available to public. In the data preprocessing phase, we proactively enhance the dataset's quality. Specifically, we reformat the dataset and address characters that pose challenges during the parsing process. It should be noted that, while we perform these operations to ensure smooth data handling, we also keep the original information in the dataset entirely unchanged. Our focus lies on rectifying technical obstacles without altering the underlying content, thus maintaining the integrity of the data.

Following the preprocessing of these passwords, the attacker uses unsupervised learning to train the representation learning model to generate honeywords for a given input password. We employed a FastText model \cite{fasttext-word-rep} as the representation learning model, as suggested in \cite{DA21}. The pre-trained LLM GPT-3.5 \cite{HoneyGAN-YF} and the heuristic-based method suggested by Dionysiou et al. \cite{DA21} are also utilized in addition to the GAN-based strategy to generate honeywords.

In Phase II (Tokenization, Model training, and Attack Phase), a CNN model that initiates with an embedding layer, where each character $\textit{c}_{i}$ of an input string \textit{s} (either a password or a honeyword) is transformed into a vector representation $\textit{v}_{i}$ using an embedding function \textit{E}. This process is mathematically represented as $s = c_1c_2 \ldots c_n \rightarrow (v_1, v_2, \ldots, v_n), \text{where } v_i = E(c_i)$. 

Following the embedding layer, the architecture includes two distinct block types that are repeated multiple times. Block 1, which repeats five times, consists of a convolutional layer for feature extraction, followed by batch normalization to stabilize and accelerate the learning process, an activation function to introduce non-linearity, pooling to reduce spatial dimensions, and dropout to mitigate overfitting by randomly omitting units used during training. The operations in Block 1 can be sequentially described by the equation $x_l = D_l(P_l(\textit{ActFn}(B_l(C_l(x_{l-1})))))$, where $C_l$ represents the convolution, $B_l$ denotes batch normalization, \textit{ActFn} denotes activation function, $P_l$ indicates pooling, and $D_l$ stands for dropout. The input $x_0$ is to the first Block 1 is the output from the embedding layer.

Block 2, repeated twice, includes a dense layer that performs high-level reasoning from features extracted previously. This layer is followed by batch normalization, an activation function, and dropout to enhance model generalization. Each iteration of Block 2 can be expressed as $z_j = D_j(\textit{ActFn}(B_j(W_j y_j + b_j)))$, where $W_j$ and $b_j$ are the weight matrix and bias vector of \textit{j-th} dense layer, respectively, and $y_j$ is the input from the previous block. The final output of the model is a probability distribution over the two classes—indicating whether the input is a honeyword or a password. This probability distribution is derived using a softmax function applied to the output of the last dense layer, represented as $p = \text{softmax}(z)$. 

\begin{figure}[h!]
  \centering
  \includegraphics[width=0.7\columnwidth]
  {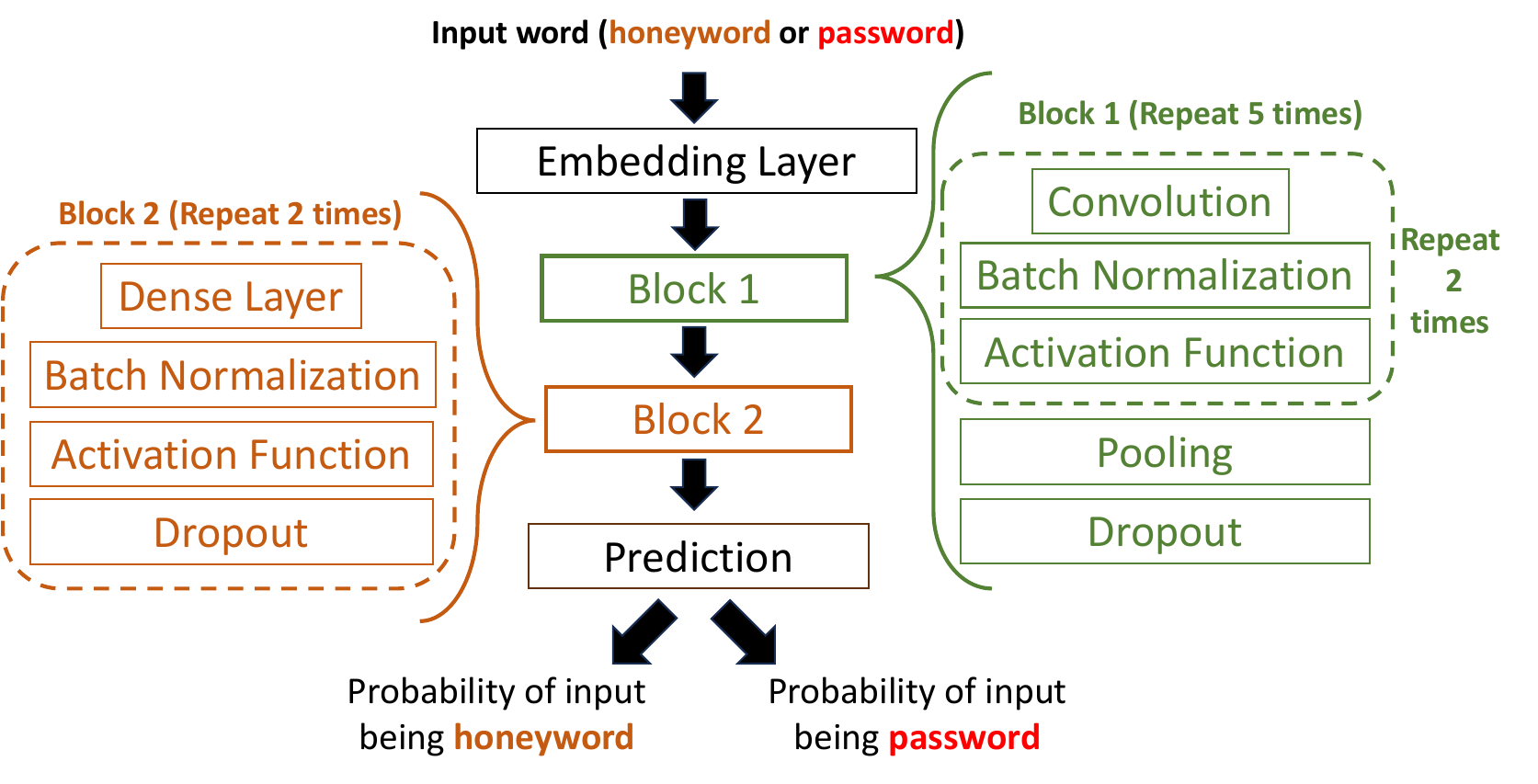}
  \caption{\textit{The architecture of CNN model used for \attackname.}}
  \label{fig:cnn-architecture}
\end{figure}

Figure~\ref{fig:cnn-architecture} shows architecture of the CNN model employed in this study. When an attacker compromises a web service using a honeywords defense mechanism and acquires the set of sweetwords, they can utilize this trained CNN model to identify passwords from the set of sweetwords associated with the user account.


We leverage this trained CNN model to compute $\epsilon$-flatness, conducting a sophisticated analysis of sweetwords' probability scores for password identification. For computing $\epsilon$-flatness, consider a collection of sweetwords $\mathcal{X}_{i}$ associated with the user $\mathcal{U}_{i}$. Each sweetword $\mathcal{X}_{ik}$ (1 $\leq$ \textit{k} $\leq$ 20) is provided as input to the classifier, and obtained the likelihood (probability score) of each of a given $\mathcal{X}_{ik}$ is password. After obtaining probability scores, rank all sweetwords related to a user account in descending order of these scores. A higher ranking or probability score suggests a greater likelihood of the sweetword being the real password. This process simulates an attacker's strategy of prioritizing guesses based on the likelihood of each sweetword being the actual password. This ranked order reflects the sequence in which an attacker, using the model's output as a guide, would likely attempt to guess the real password. 

Additionally, to identify suspicious activity the system should alert administrators of suspicious login attempts with honeywords, triggering an alarm if login attempts exceed a set threshold $\mathcal{TH}_{1}$ for a user account. A system-wide alarm activates if attempts surpass a broader threshold $\mathcal{TH}_{2}$. These thresholds are vital for safeguarding against malicious access and are set based on the risk assessment of the targeted system \footnote{Determining precise threshold values is beyond the scope of this study.}. The effectiveness of an attack is measured by the attacker's success within the permissible number of attempts, as determined by these thresholds.

%% file: results.tex
\section{Evaluation Framework}
\label{sec:eval}
\noindent
In this section, we define the framework to evaluate \attackname\ in identifying passwords from the set of sweetwords. We describe the datasets, experiment settings, and evaluation metric considered for evaluating our attack. 

\subsection{Description of the Datasets}
\label{sub-sec:describe-datasets}
\noindent
In our study, we assessed the efficacy of the \attackname\ attack using publicly accessible datasets made available by Dionusiou et al. Similar datasets have been instrumental in prior studies, highlighting the significance of consistent data sources for advancing research. For example, studies \cite{BH17, DA21, DE14, MJ14, WM09SM} used identical datasets to explore related research topics. These datasets feature passwords from various web services listed in Table~\ref{tab:count-of-available-passwords}, compromised and released publicly by hackers or insiders. Dionusiou et al. preprocessed these datasets, specifically by excluding passwords that are shorter than eight characters. Moreover, the RockYou dataset was released in two versions, referred to as RockYou-large and RockYou-small in this study. 


\begin{table}[h!]
\centering
\caption{\textit{The number of passwords available for each web service from the processed passwords published in \cite{DA21}.}}
\label{tab:count-of-available-passwords}
\begin{tabular}{|c|c|}
\hline
\textbf{Dataset}                                                                                                                     & \textbf{No. of Passwords} \\ \hline
\begin{tabular}[c]{@{}c@{}}Adultfriendfinder, Dubsmash, DropBox, Myspace, Phpbb\\ RockYou-small, LinkedIn, Yahoo, Zynga\end{tabular} & 50,000 for each           \\ \hline
RockYou-large                                                                                                                        & 1.04M                     \\ \hline
\end{tabular}
\end{table}

The honeywords utilized for training the classifiers were generated using the implementation provided by Dionysiou et al. Our approach involved creating a one-to-one correspondence between passwords and honeywords to train the model effectively. For each given password \textit{p}, we selected the honeyword \textit{h} that exhibited the highest cosine similarity to \textit{p}, as generated by the FastText model. This strategy ensured that each password had a unique honeyword, allowing the CNN model to learn a balanced and optimized set of passwords and honeywords that were strongly correlated.

Furthermore, to generate honeywords using the Honey-Chunk \cite{HoneyGAN-YF} approach, which employs a pre-trained LLM GPT-3.5 (developed by OpenAI \cite{chatgpt-paper}). We designed a crawler to interact with the GPT model, producing a diverse set of honeywords directly from the corresponding password data. This approach aimed to enhance the dataset's quality and authenticity, specifically addressing our research needs. 

\noindent
\textbf{Data representation:} To facilitate the learning process of the CNN model, we employed tokenization where passwords and honeywords are converted into character-level representations. Each character within a password or honeyword was considered an individual token. Given the inherent variability in the length of passwords and honeywords, padding was applied to standardize the length of these character-level tokens. This standardization process is imperative for the training phase of the CNN model, as it necessitates input of a consistent size.

\noindent
\textbf{Ethical considerations:} In our study, we follow the ethical framework proposed by Thomas et al. \cite{DT17}, we exclusively used public datasets containing only passwords, ensuring there was no additional PII, such as email addresses, involved. This selection was imperative as our research relies on analyzing password datasets without causing any undue harm or privacy breaches; specifically, we ensured that the passwords under study could not be traced back to individual PII. Furthermore, our aim is not to identify users from these leaked passwords but rather to leverage them for generating honeywords and for training deep learning models.

\subsection{Experiment Settings}
\label{sub-sec:experiment-settings}
\noindent
In this study, the implementation of the \attackname\ attack was performed using two open-source libraries: \texttt{FastText} \cite{fasttext-word-rep} and \texttt{TensorFlow} \cite{tensorflow2015-whitepaper} for training DNNs.  Utilizing \texttt{FastText} to train the representation learning model and \texttt{TensorFlow} to train and evaluate the CNN model, we developed a robust \attackname\ attack that delivered state-of-the-art performance in identifying passwords from the set of sweetwords associated with the user account.

\noindent
\textbf{FastText training:} Following Dionysiou et al.’s guidelines, we trained the FastText model on RockYou-large dataset as the size of dataset has substantial impact of model training. It is crucial to note that insufficient data during model training can lead to various issues, such as increased loss. This limitation might cause the FastText model to generate less reliable, contextually inappropriate, or lower-quality honeywords. For a comprehensive understanding of the hyper-parameters, we direct the readers to \cite{DA21}.

\noindent
\textbf{CNN training:} We used the CNN model to distinguish between passwords and honeywords linked to a user account. The CNN model's architecture is adopted from the architecture presented in \cite{SPIM18, JD21}. In our study, we replaced the input layer of the model with an embedding layer because the passwords and honeywords that are represented as character-level tokens should be converted to numeric representation. The Embedding layer transforms the characters into a dense vector representation, beneficial for multiple tasks in the NLP domain \cite{PBLL04, SMNK20}. Moreover, we employed Microsoft Neural Network Intelligence (Microsoft NNI) \cite{microsoft-nni} to find optimal hyperparameters as the choice of hyperparameters significantly impacts the model's classification performance. For each of the datasets described in Table~\ref{tab:count-of-available-passwords}, we selected 90,000 (45,000 passwords, and 45,000 corresponding honeywords) data points that were divided into training (90\%), validation (5\%), and testing (5\%). We conducted validation and testing by dividing the data into categories, which prevented underfitting and overfitting of the deep learning model. Finally, to rigorously assess the generalizability of \attackname\ on unseen sweetwords, we conducted evaluations using an independent dataset derived from the remaining 5,000 passwords of the original 50,000-password pool, after allocating 45,000 for training, validation, and testing. This subset, comprising 20 sweetwords per account for a total of 500 user accounts, was distinct from those used in earlier phases, ensuring that the model’s performance was evaluated in an unbiased and reliable manner.

\subsection{\attackname\ Evaluation Metric: $\epsilon$-flatness}
\label{sub-sec:evaluation-metric}

\noindent
We evaluate the effectiveness of our approach in guessing passwords in \textit{x} attempts by establishing a baseline comparison with the random guessing on chosen datasets. To maintain consistency with previous studies \cite{DA21}, we replicate the methodology employed by Dionysiou et al. for honeywords generation. To thoroughly assess the effectiveness of our approach, we utilized an independent dataset tailored for this purpose. This dataset consisted of passwords obtained from 500 users, and for each password, we generated 19 corresponding honeywords, resulting in a total of 20 sweetwords associated with each user account (as suggested by Juels and Rivest \cite{JA13}). 

``$\epsilon$-flatness'' measures the highest probability that an adversary can correctly guess the real password from the set of sweetwords. For example, if a attack achieves $\epsilon$-flatness with $\epsilon$ = 0.05, an adversary’s chances of correctly guessing the real password are limited to 5\%. We applied the $\epsilon$-flatness metric to critically assess the efficacy of our password guessing attack in our evaluation of \attackname. However, the original metric proposed by Juels and Rivest \cite{JA13} did not support scenarios where an attacker makes multiple guesses per user account, nor did it highlight the system’s most vulnerable honeywords. Wang et al. updated this metric to accommodate these scenarios, enabling it to handle multiple login attempts by an attacker and more accurately represent the effectiveness of HGTs in producing secure honeywords.

The $\epsilon$-flatness metric can be visualized using a flatness graph, where a point ($\textit{x}$, $\textit{y}$) illustrates that within the first $\textit{x}$ attempts, the likelihood of correctly identifying the real password is $\textit{y}$, as long as $\textit{x}$ $\leq$ $\textit{k}$. Here, $\textit{k}$ represents the number of sweetwords per user account \cite{DW18}. In reality, the system employing honeywords considers threshold ($\mathcal{TH}_{1}$). Surpassing $\mathcal{TH}_{1}$ failed login attempts alerts the system administrator, indicating a possible data breach. This approach carefully balances the need to provide legitimate users sufficient attempts while rigorously monitoring for unauthorized access. For further insights into the process of obtaining e-flatness by analyzing sweetwords as passwords, see Section~\ref{sec:methodology}. To generate the flatness graph and quantify the performance of the model using \textit{x} guessing attempts two principal axis are considered:

\begin{enumerate}[leftmargin=0.5cm]
    \item \textit{X-axis (Number of Attempts):} This axis quantifies the number of \textit{x} guessing attempts an attacker undertakes to identify the correct password.
    \item \textit{Y-axis (Probability of Correct Guess):} This dimension measures the likelihood of an attacker successfully guessing the real password within a \textit{x} number of attempts, as specified on the X-axis. The probability of a correct guess for each point on the X-axis is calculated by determining the proportion of instances where the genuine password is ranked within the top-\textit{x} guesses, based on the model's probability scores.
\end{enumerate}

In our research, the average e-flatness across multiple datasets under various experimental settings for a group of 500 users is considered for evaluating the overall performance of the attack. We have designated this average e-flatness metric as the `success rate' of \attackname\ for identifying passwords.

\section{Evaluation \& Results}
\label{sub-sec:attack-experiments}
\noindent
In this section, we evaluate \attackname\ across various threat models outlined in this paper and compare its performance against state-of-the-art HGTs. 

\smallskip
\noindent 
\textbf{Experiment A. Attack evaluation with threat model 1 (same-service model).} In this experiment, we evaluate \attackname\ in same-service threat model where an attacker repeatedly compromises same web service. The attacker initially compromises the honey checker along with the web service, resulting in attacker obtaining labeled dataset that is used to train the CNN model. In subsequent attacks to that web service, the attacker targets only the web server that stores the sweetwords and steals the file containing them. To ensure consistency and evaluate the efficacy of \attackname\ with multiple web services, we trained and evaluate the CNN model on the dataset from each web service independently. Meaning, if the CNN model was trained on the dataset obtained from ``\textit{Dubsmash.com}'' web service, the evaluation will use a different set of sweetwords from the same web service (i.e., ``\textit{Dubsmash.com}'').

The average success rate across 500 users when an attacker can successfully log in to the system with first attempt of the sweetword identified as password by \attackname\ with highest probability, is significantly higher as compared to random guessing. The results show that when considering chaffing-by-tweaking HGT, the attacker's success rate of identifying passwords is 52.78\%, which is 10.556$\times$ more effective than the 5\% success rate achieved by random guessing. When chaffing-with-a-hybrid-model HGT is considered, \attackname\ achieves a success rate of 51.62\% which is 10.324$\times$ higher as compared to random guessing. However, the effectiveness of \attackname\ drops significantly to 14.82\% when chaffing-with-a-password-model HGT is considered, this success rate is 2.964$\times$ higher than that achieved by random guessing. Despite the chaffing-with-a-password-model generating more realistic honeywords, \attackname\ significantly outperforms the random guessing success rate.

To further quantify the efficacy of the attack when three login attempts are permitted, the average success rates for different HGTs show substantial improvements. Specifically, the success rate increases to 79.22\%, 35.76\%, and 80.13\% for chaffing-by-tweaking, chaffing-with-a-password-model, and chaffing-with-a-hybrid-model, which is 5.28$\times$, 2.38$\times$, and 5.34$\times$, respectively, improvement as compared to 15\% random guessing password identification success rate. By the time ten attempts are made, the success rate improves significantly by 98.18\%, 80.38\%, and 98.93\% for chaffing-by-tweaking, chaffing-with-a-password-model, and chaffing-with-a-hybrid-model, which is 1.96$\times$, 1.60$\times$, and 1.97$\times$ improvement respectively, as compared to 50\% random guessing. However, despite its proximity to random guessing, which indicates enhanced security, the chaffing-with-a-password-model remains susceptible. This shows that the attack's average success rate using this model still surpasses the expected rate of chance agreement across all attempts, highlighting all the HGTs are vulnerable.

\begin{figure*}[htp!]
     \centering
     \begin{subfigure}[t]{0.35\textwidth}
         \centering
         \includegraphics[width=\textwidth]{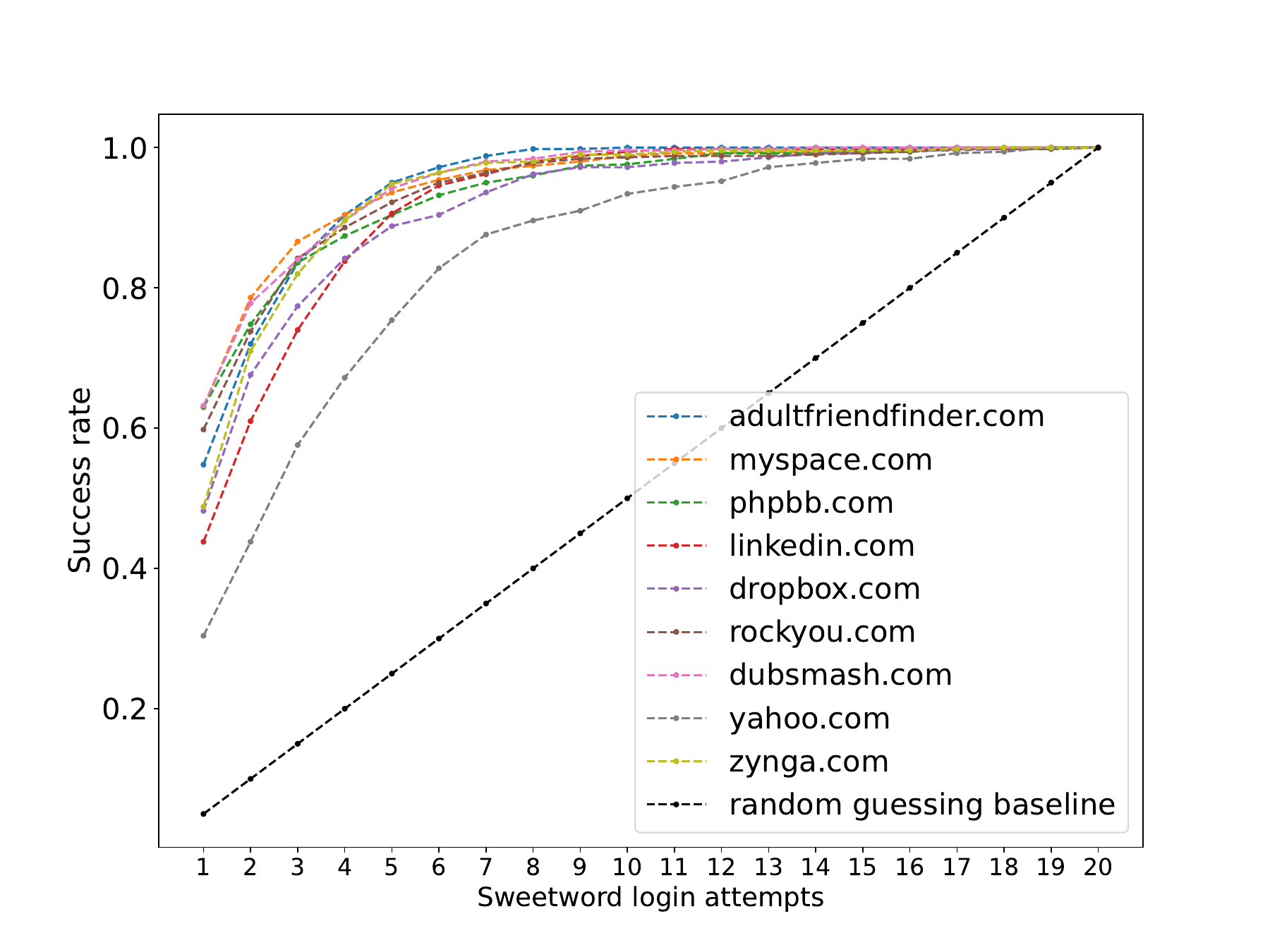}
         \caption{Chaffing-by-tweaking}
         \label{fig-exp-ss-tweaking}
     \end{subfigure}
     \hspace{-7mm} 
     \begin{subfigure}[t]{0.35\textwidth}
         \centering
         \includegraphics[width=\textwidth]{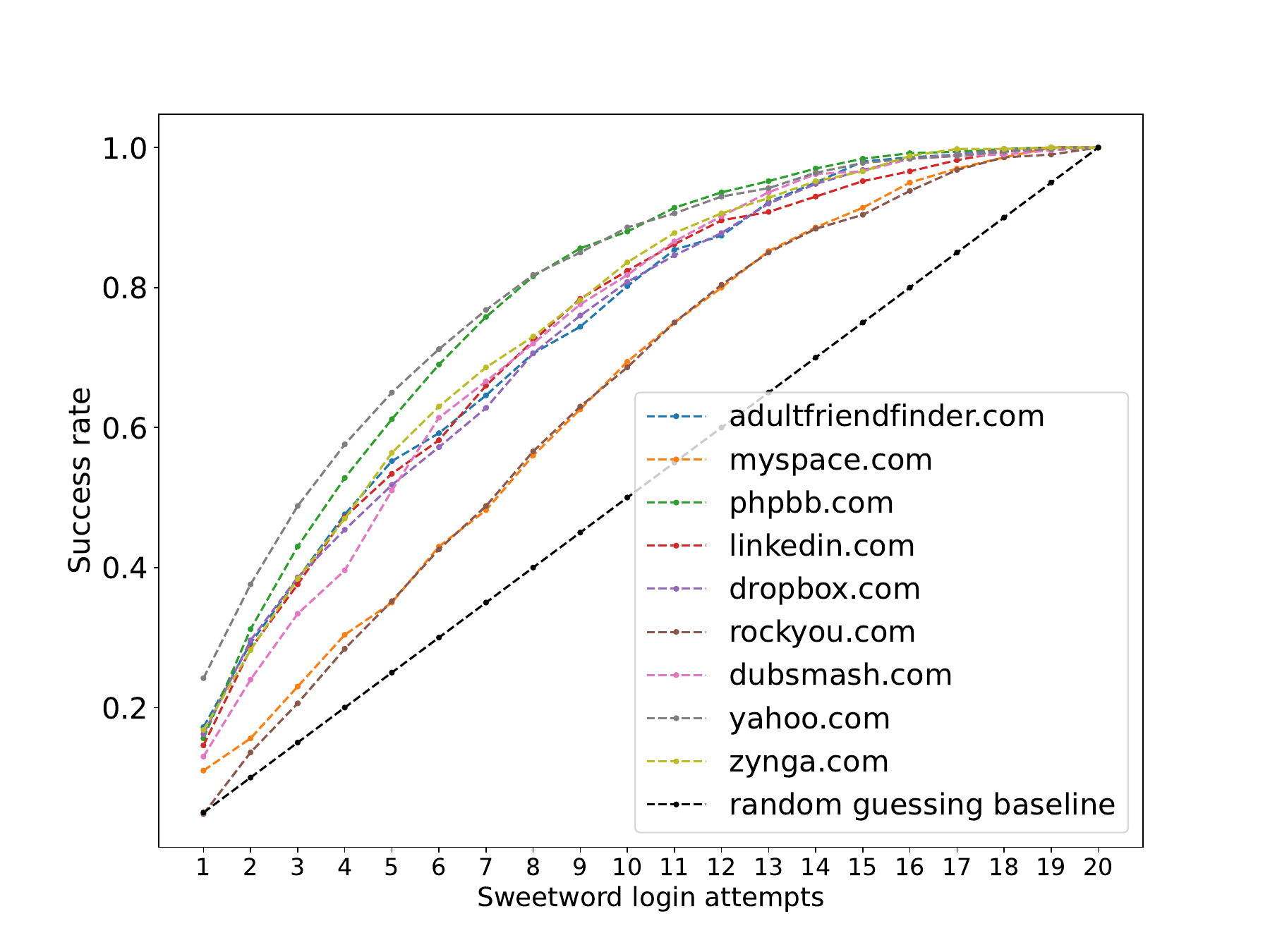}
         \caption{Chaffing-with-a-password-model}
         \label{fig-exp-ss-model}
     \end{subfigure}
     \hspace{-7mm} 
     \begin{subfigure}[t]{0.35\textwidth}
         \centering
         \includegraphics[width=\textwidth]{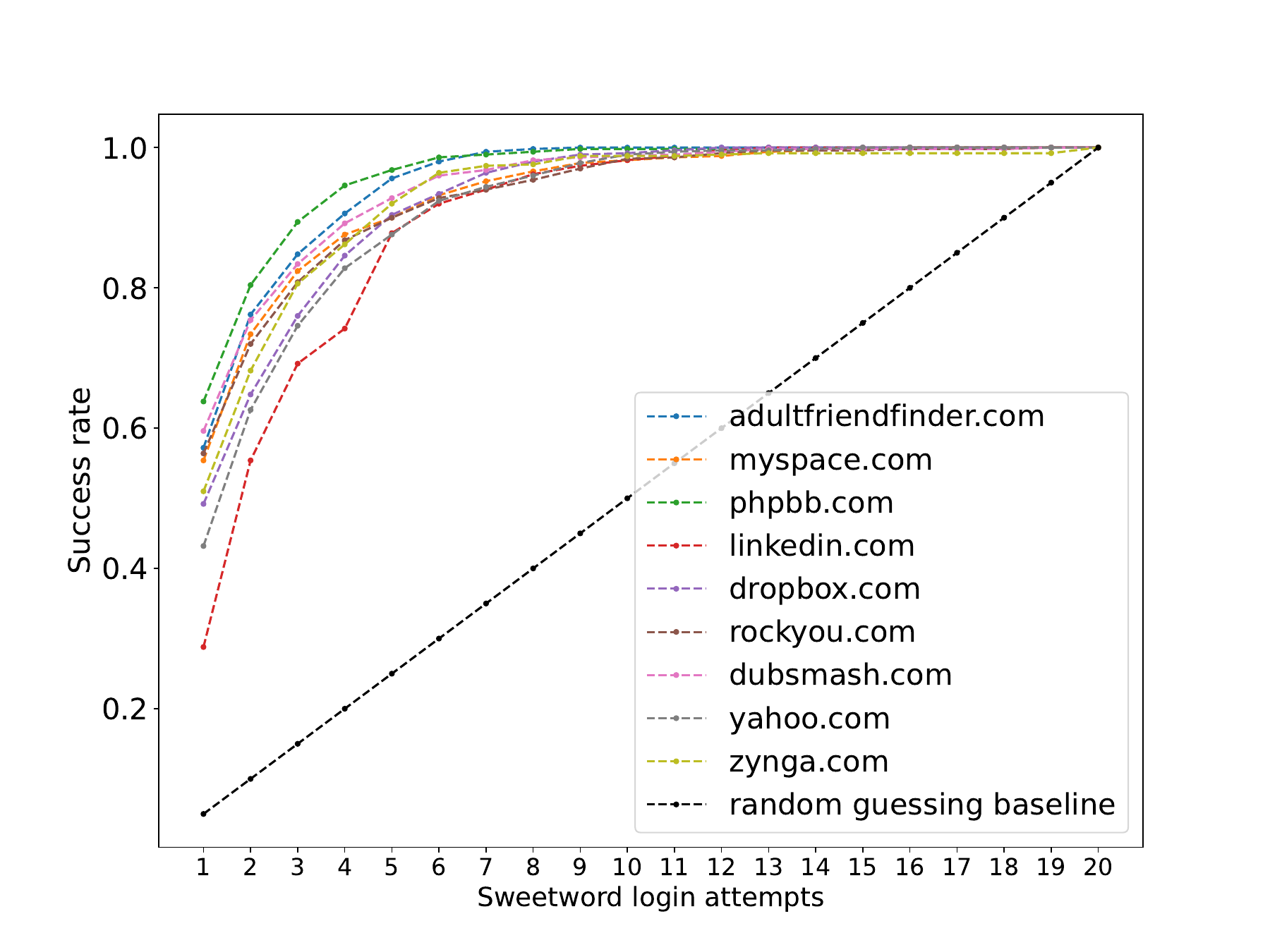}
         \caption{Chaffing-with-a-hybrid-model}
         \label{fig-exp-ss-hybrid-model}
     \end{subfigure}
        \caption{\textit{The flatness graphs of \attackname\ when \textbf{same-service threat model} is considered as an attack model. The closer to random guessing baseline, the better security is provided by the corresponding HGT.}}
        \label{fig-exp-ss}
\end{figure*}

\begin{figure}[h!]
  \centering
  \includegraphics[width=0.7\textwidth]{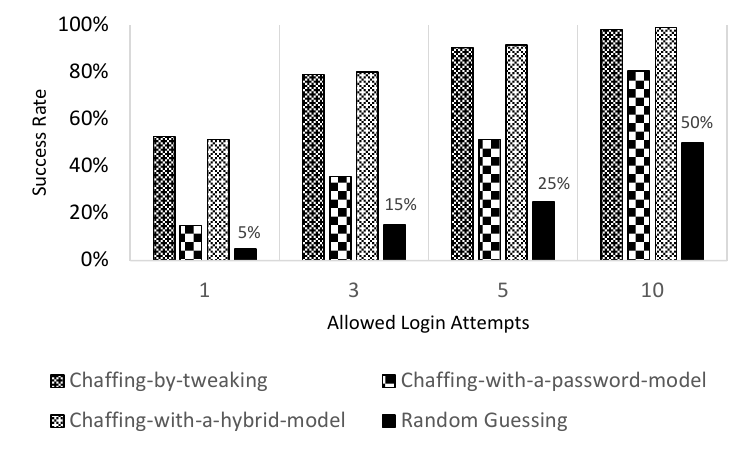}
  \caption{\textit{Summary of \attackname\ on different HGTs in \textbf{same-service threat model} when number of allowed login attempts are 1, 3, 5, and 10.}}
  \label{fig:summary-of-results-exp-a}
\end{figure}

Moreover, the flatness graphs for each HGT depicted in Figures~\ref{fig-exp-ss-tweaking}, ~\ref{fig-exp-ss-model}, and ~\ref{fig-exp-ss-hybrid-model}, illustrate that the attack's performance significantly enhances with increasing number of login attempts. These graphs reveal that the performance curves for different datasets are closer to the random guessing line when employing the chaffing-with-a-password-model HGT, in contrast to the curves generated by chaffing-by-tweaking and chaffing-with-a-hybrid-model. Figure~\ref{fig:summary-of-results-exp-a} summarized the performance of \attackname\ with increasing login attempts. The $\epsilon$-flatness of \attackname\ on individual dataset with 1, 3, 5, and 10 attempts is presented Table~\ref{tab:exp-a-1-3-5-10-attempts} (Appendix~\ref{lab:exp-3-5-10-attempts}).

\smallskip
\noindent
\textbf{Experiment B. Attack evaluation with threat model 2 (cross-service model).} In this experiment, we assess \attackname\ using a cross-service threat model. In this scenario, an attacker compromises one web service and extracts passwords and honeywords to train a CNN model. This trained model is then deployed to identify real passwords from a collection of sweetwords from another compromised web service. For instance, suppose the attacker initially trains the CNN model with passwords and honeywords from the ``\textit{Dubsmash.com}'' web service. If the attacker later compromises another service, such as ``\textit{MySpace.com}'', the attacker can then apply the model trained on ``\textit{Dubsmash.com}'' to discern the real passwords from the sweetwords associated with ``\textit{MySpace.com}''. To simulate this attack model, we trained a CNN model on passwords and honeywords of the ``\textit{Dubsmash.com}'' and utilized that model to identify passwords for other eight web services.

In cross-service settings, when the attacker logs in to the system on the first attempt using the sweetword identified as the most likely password, PassFilter consistently outperforms random guessing. However, the average success rate in cross-service threat models is slightly lower than in same-service settings. This variation is attributed to differences in the distribution of passwords across various websites. When the CNN model is evaluated using data from a different web service than the one it was trained on, there is a noticeable drop in performance. Specifically, the average success rates are 51.84\%, 8.22\%, and 48.13\% for chaffing-by-tweaking, chaffing-with-a-password-model, and chaffing-with-a-hybrid-model HGTs, respectively. These rates are significantly better—10.36$\times$, 1.64$\times$, and 9.626$\times$, respectively—compared to the 5\% success rate achieved by random guessing with 20 sweetwords per account.

\begin{figure*}[htp!]
     \centering
     \begin{subfigure}[b]{0.35\textwidth}
         \centering
         \includegraphics[width=\textwidth]{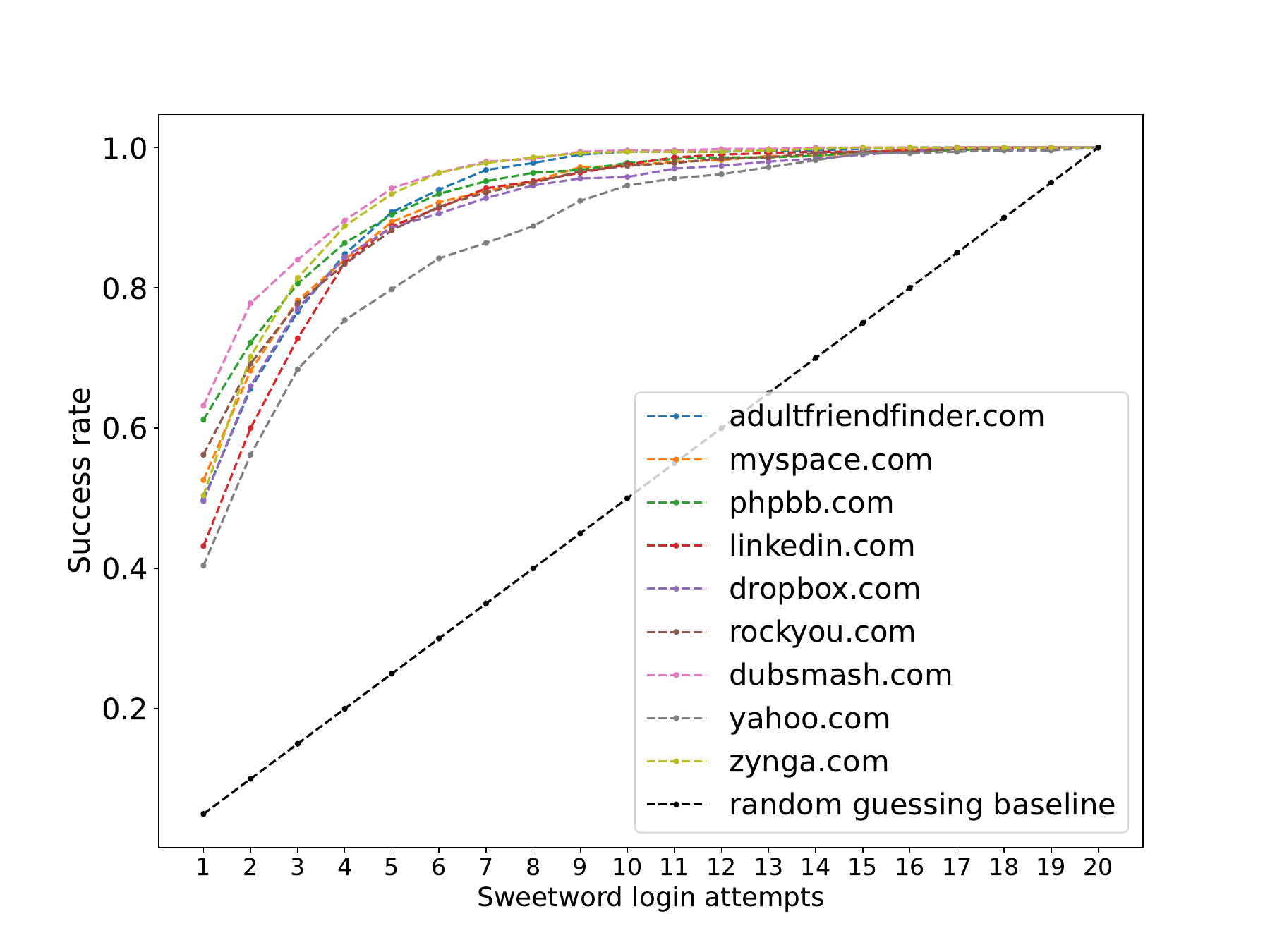}
         \caption{Chaffing-by-tweaking}
         \label{fig-exp-cs-tweaking}
     \end{subfigure}
     \hspace{-7mm}
     \begin{subfigure}[b]{0.35\textwidth}
         \centering
         \includegraphics[width=\textwidth]{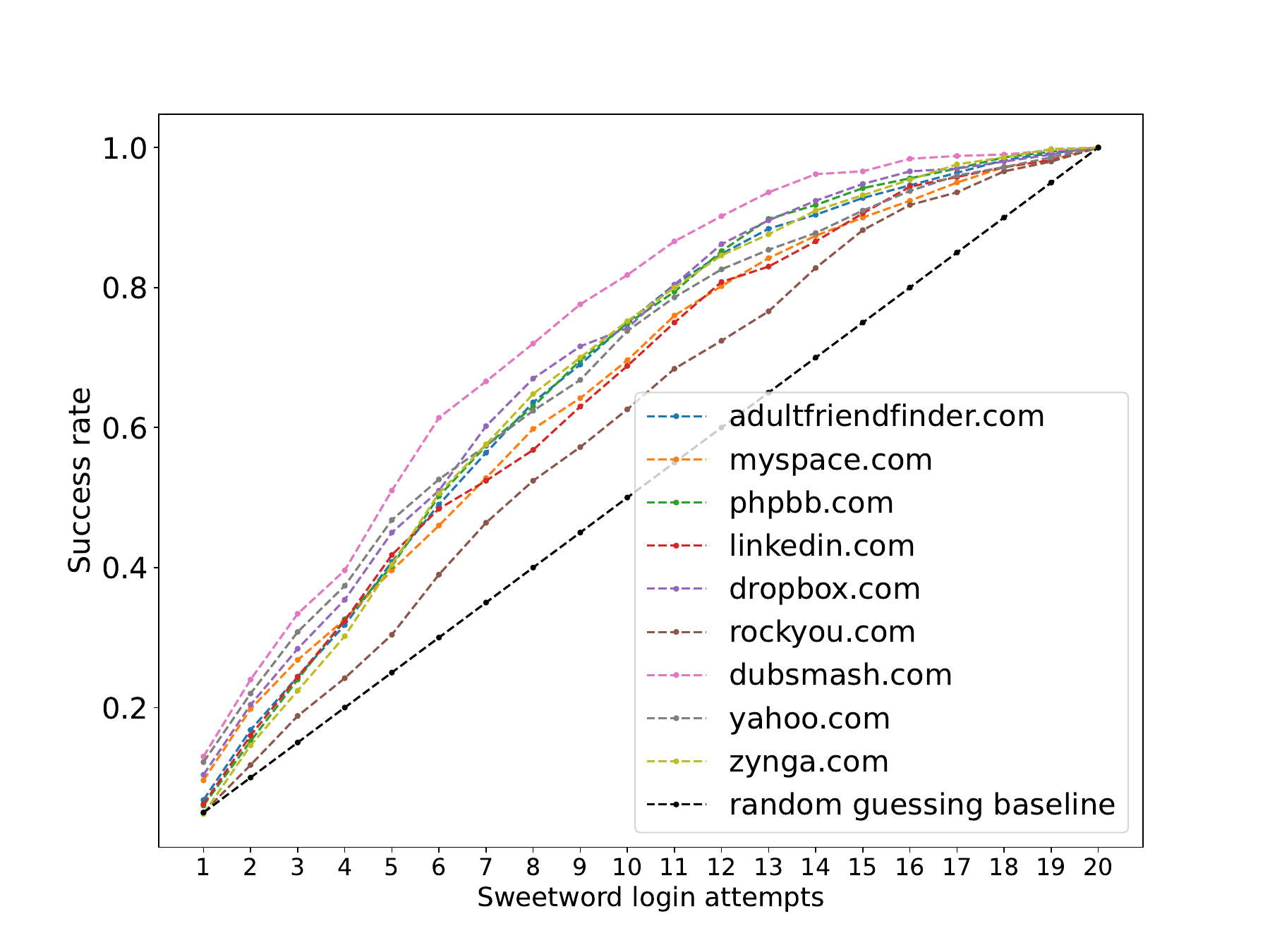}
         \caption{Chaffing-with-a-password-model}
         \label{fig-exp-cs-model}
     \end{subfigure}
     \hspace{-7mm}
     \begin{subfigure}[b]{0.35\textwidth}
         \centering
         \includegraphics[width=\textwidth]{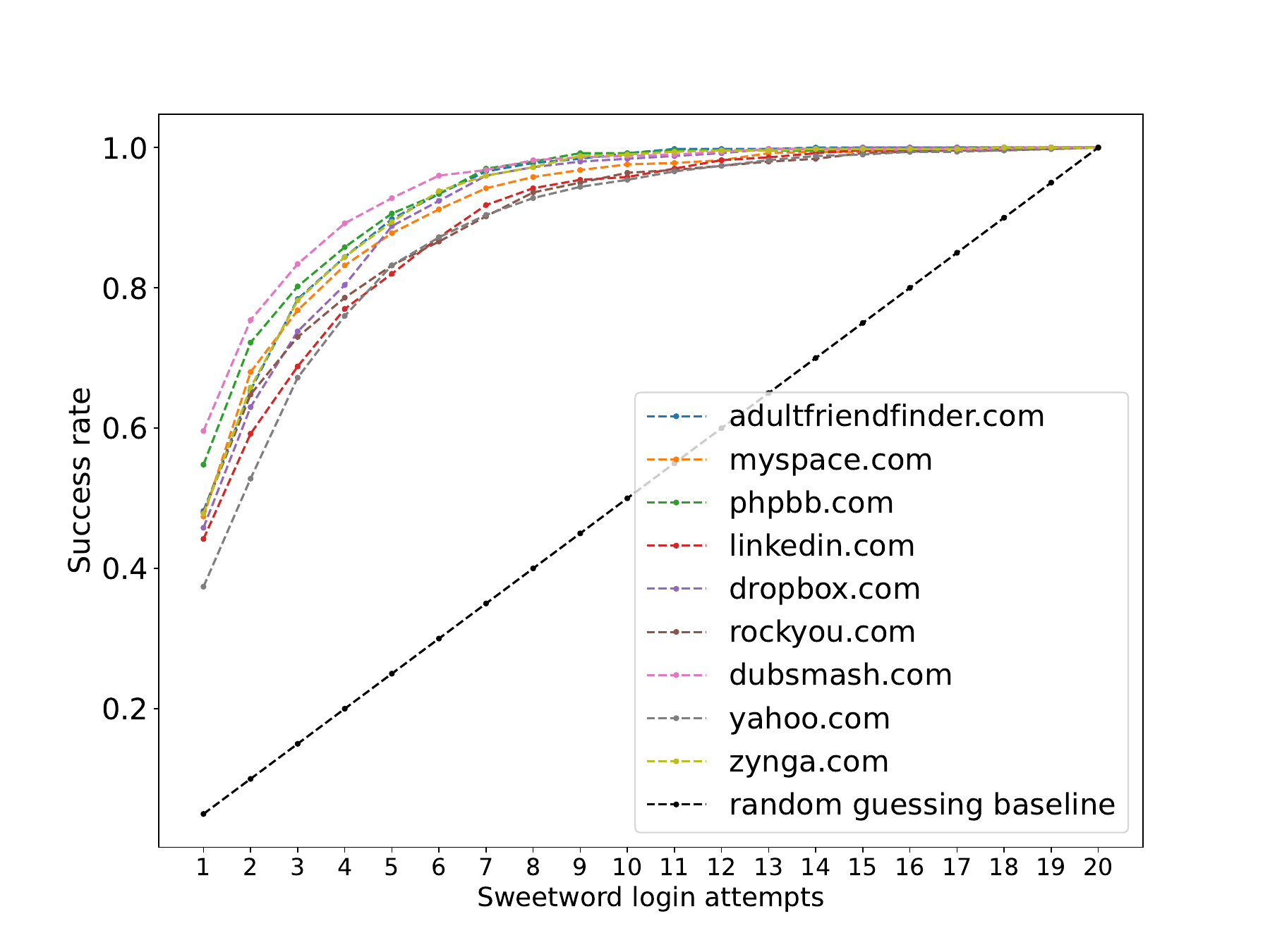}
         \caption{Chaffing-with-a-hybrid-model}
         \label{fig-exp-cs-hybrid-model}
     \end{subfigure}
        \caption{\textit{The flatness graphs of \attackname\ when \textbf{cross-service threat model} is considered as an attack model. The closer to random guessing baseline, the better security if provided by the corresponding HGT.}}
        \label{fig-exp-cs}
\end{figure*}

\begin{figure}[h!]
  \centering
  \includegraphics[width=0.7\textwidth]{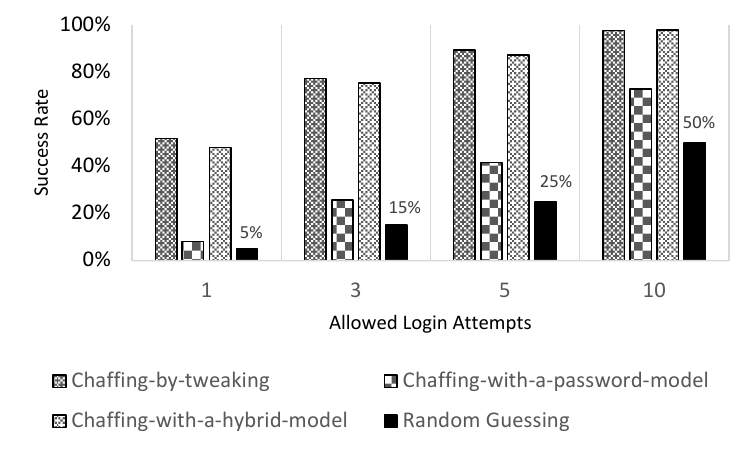}
  \caption{\textit{Summary of \attackname\ on different HGTs in \textbf{cross-service threat model} when number of allowed login attempts are 1, 3, 5, and 10.}}
  \label{fig:summary-of-results-exp-b}
\end{figure}

Furthermore, with increasing login attempts, we observed significant improvement in password identification. With just three login attempts, the chaffing-by-tweaking, chaffing-with-a-password-model, and chaffing-with-a-hybrid-model approaches showed average success rates of 77.42\%, 25.93\%, and 75.53\%, which is 5.16$\times$, 1.72$\times$, and 5.03$\times$ better as compared to random guessing (15\%), respectively. Similarly, for ten login attempts, the success rates of \attackname\ increase to 97.67\%, 72.87\%, and 97.76\% for chaffing-by-tweaking, chaffing-with-a-password-model, and chaffing-with-a-hybrid-model, respectively. This shows the improvement by the factor of 1.95$\times$ for chaffing-by-tweaking, and chaffing-with-a-hybrid-model HGTs, and 1.45$\times$ improvement for chaffing-with-a-password-model. The success rate of \attackname\ with increasing number of login attempts for individual dataset for cross-service settings is presented in Table~\ref{tab:exp-b-1-3-5-10-attempts} in Appendix~\ref{lab:exp-3-5-10-attempts}. This highlights the robustness of our attack in password identification, consistently demonstrating a higher likelihood of evading honeywords defense mechanism. These performance enhancements with increasing number of login attempts are summarized in Figure~\ref{fig:summary-of-results-exp-b}.

The performance of \attackname\ in cross-service scenarios is depicted in the flatness graphs in Figure ~\ref{fig-exp-cs-tweaking}, ~\ref{fig-exp-cs-model}, and ~\ref{fig-exp-cs-hybrid-model} for all examined HGTs. These graphs show that the chaffing-with-a-password-model HGT consistently maintains low success (but higher as compared to a chance agreement) on the first attempt, proving its efficacy in creating realistic honeywords. Conversely, the chaffing-by-tweaking and chaffing-with-a-hybrid-model techniques exhibit higher success rates, allowing password identification with fewer attempts.

\smallskip 
\noindent
\textbf{Experiment C. Attack evaluation with threat model 3 (self-trained model).} In this experiment, we evaluate the effectiveness of \attackname\ where the attacker gains access to a web service that already utilizes honeywords as a defense mechanism, that is, the attacker does not compromise honey checker to obtain the labeled dataset. Specifically, we evaluated self-trained threat model, where the attacker generated passwords using the PassGAN \cite{HB19} model.

\attackname\ demonstrates a significantly higher success rate in password identification in self-trained threat model across different HGTs when 20 sweetwords are associated with user accounts. In first login attempt, chaffing-by-tweaking and chaffing-with-a-hybrid-model shows the average success rate of 47.90\%, and 36.90\%, respectively. These success rates are 9.58$\times$, and 7.38$\times$ better as compared to random guessing (5\%). For chaffing-with-a-password-model, the average success rate is 6.10\% which is moderately higher as compared to random guessing. The success rate of \attackname\ with increasing number of login attempts for individual dataset for adversarial threat model is presented in Table~\ref{tab:exp-c-1-3-5-10-attempts} in Appendix~\ref{lab:exp-3-5-10-attempts}. Figure~\ref{fig:summary-of-results-exp-c} summarized the performance of \attackname\ in self-trained threat model when number of login attempts allowed are 1, 3, 5, and 10.

\begin{figure*}[htp!]
     \centering
     \begin{subfigure}[b]{0.35\textwidth}
         \centering
         \includegraphics[width=\textwidth]{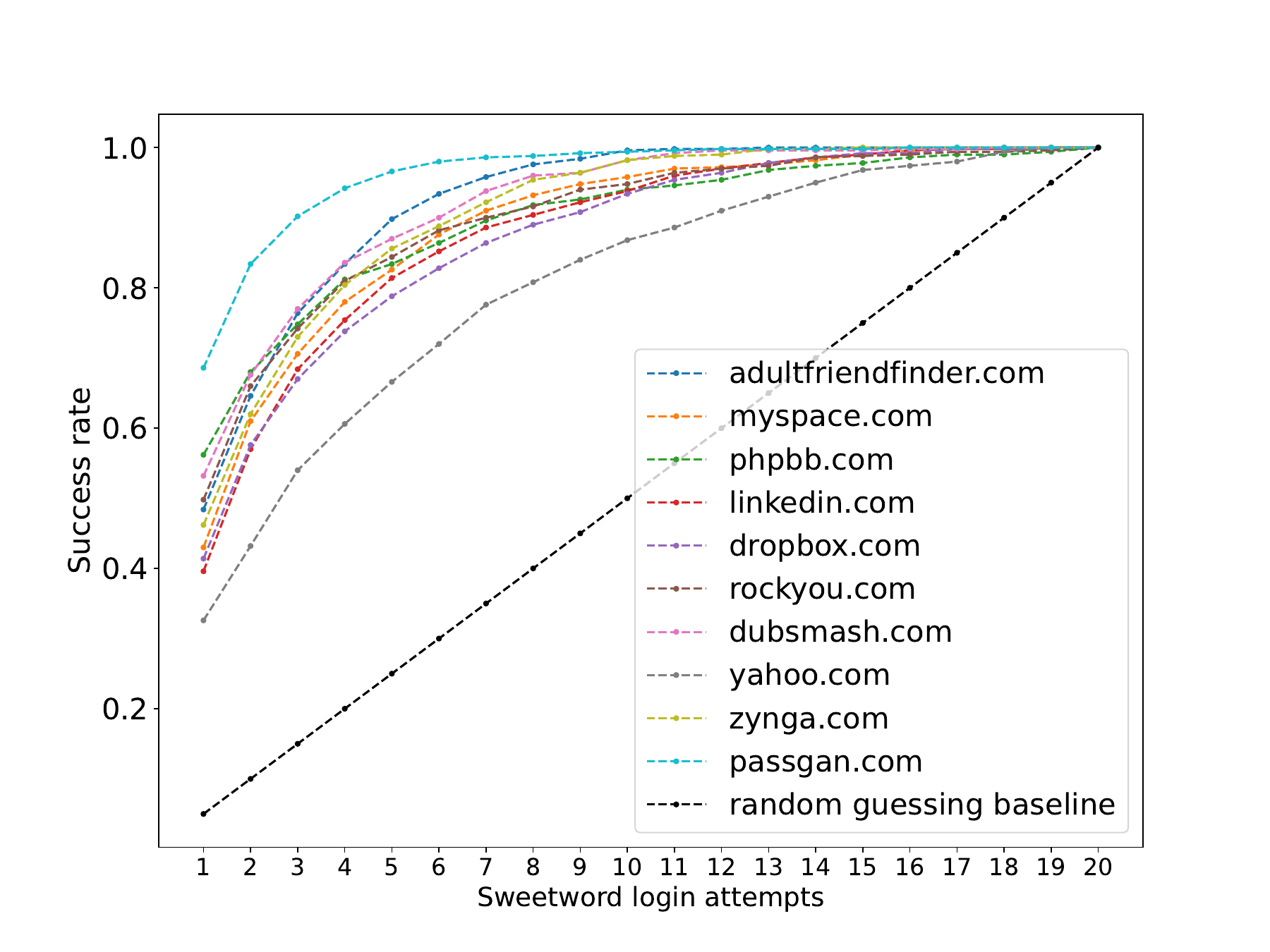}
         \caption{Chaffing-by-tweaking}
         \label{fig-exp-passgan-tweaking}
     \end{subfigure}
     \hspace{-7mm}
     \begin{subfigure}[b]{0.35\textwidth}
         \centering
         \includegraphics[width=\textwidth]{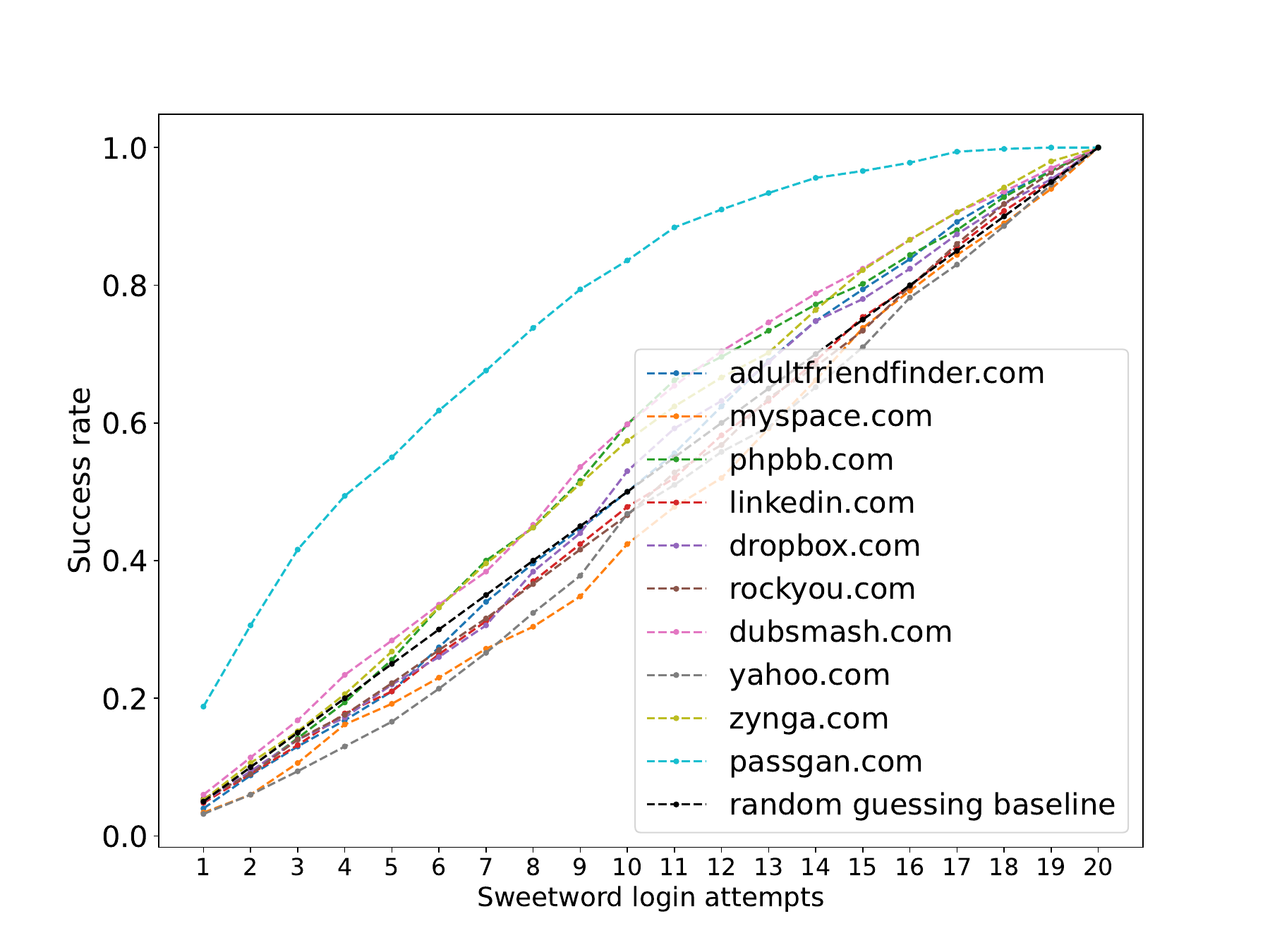}
         \caption{Chaffing-with-a-password-model}
         \label{fig-exp-passgan-model}
     \end{subfigure}
     \hspace{-7mm}
     \begin{subfigure}[b]{0.35\textwidth}
         \centering
         \includegraphics[width=\textwidth]{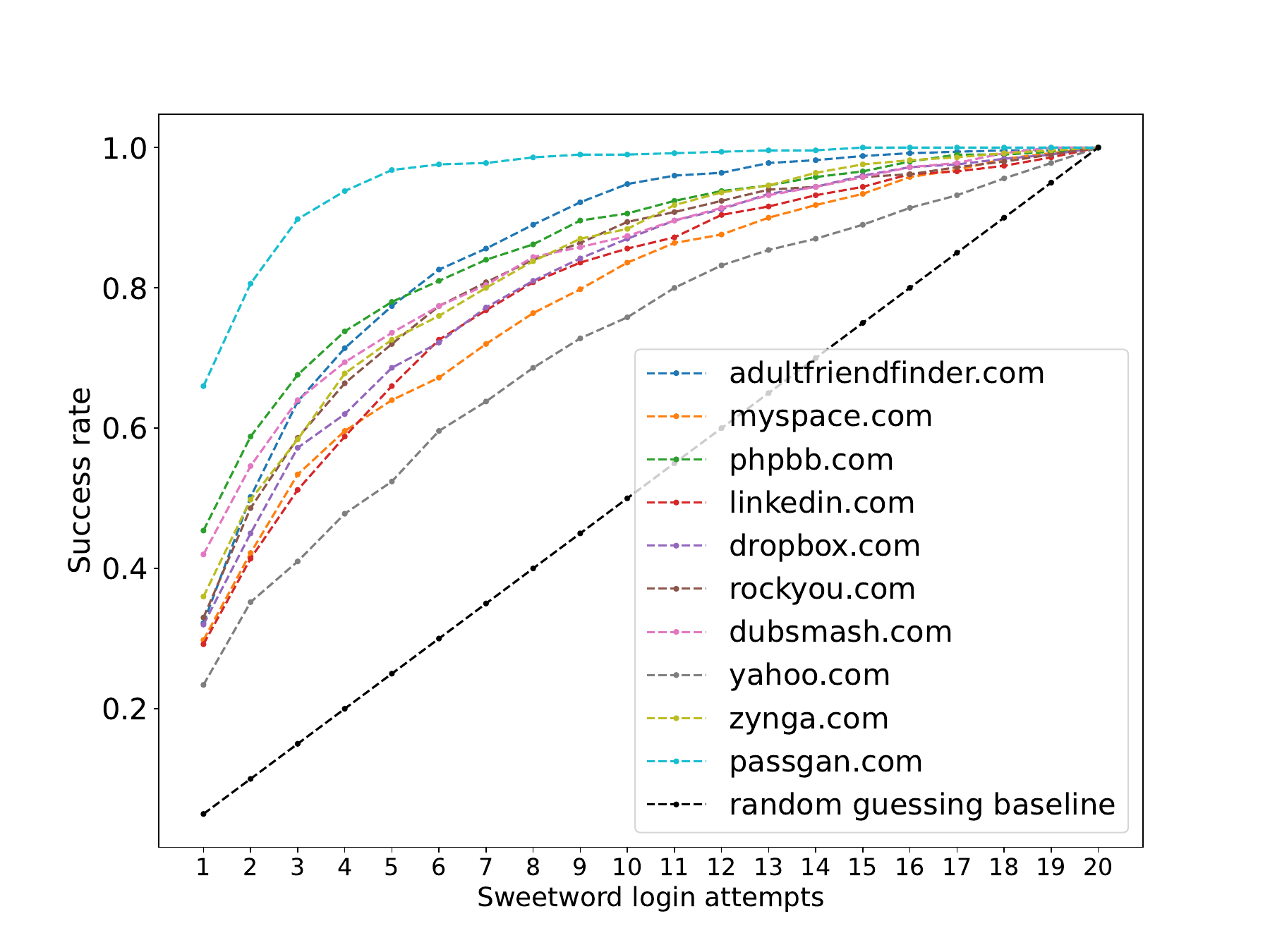}
         \caption{Chaffing-with-a-hybrid-model}
         \label{fig-exp-passgan-hybrid-model}
     \end{subfigure}
        \caption{\textit{The flatness graphs of \attackname\ when \textbf{self-trained threat model} is considered as an attack model. The closer to random guessing baseline, the better security if provided by the corresponding HGT.}}
        \label{fig-exp-passgan}
        \vspace{-1mm}
\end{figure*}

The performance of the CNN model closely approaches random guessing for some datasets within the self-trained threat model, significantly underperforming compared to results in same-service and cross-service threat models. This decline is due to training with adversarial passwords generated by PassGAN. These passwords mimics the statistical properties of its training data, blurring the distinctions between real and adversarially generated passwords which indicates a substantial difference in the distributions of adversarially generated and real passwords, worsening CNN's difficulty in recognizing human-generated passwords.

Moreover, training the CNN on adversarially generated passwords introduces a bias where the model prioritizes adversarial cues over broader characteristics typical of genuine passwords leading to degraded performance when tested against real-world datasets lacking these cues. Hatji et al. \cite{HB19} also quantified that the passwords generated by the suggested technique matched 34.6\% of the passwords in the RockYou dataset testing set and 34.2\% of the passwords in the LinkedIn dataset. This suggests that adversarially generated passwords often contain subtle patterns that are rare in user-created passwords, complicating CNN's task of effectively identifying passwords from honeywords.

\begin{figure}[h!]
  \centering
  \includegraphics[width=0.7\textwidth]{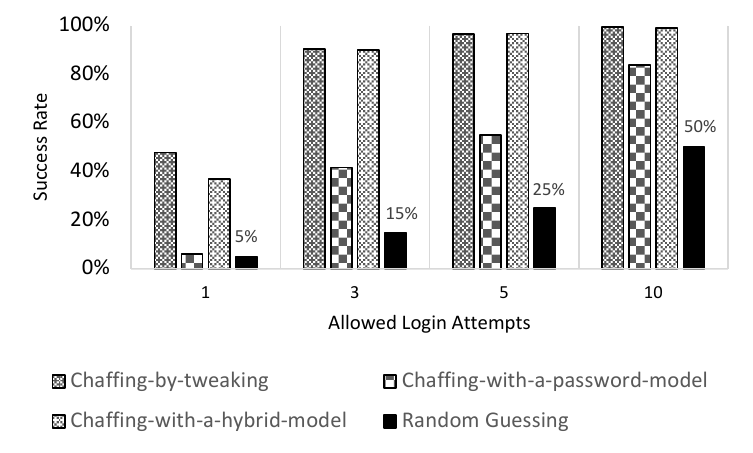}
  \caption{\textit{Summary of \attackname\ on different HGTs in \textbf{self-trained threat model} when number of allowed login attempts are 1, 3, 5, and 10.}}
  \label{fig:summary-of-results-exp-c}
\end{figure}

\noindent
\textbf{Impact of pre-trained models as HGT.} In our study, we assess the effectiveness of specialized and general-purpose language models utilized in generating honeywords. Specifically, we test the security of honeywords created by a chaffing-with-a-password model using a FastText model trained for this purpose. We also examine the impact of honeywords produced using the pre-trained general-purpose LLM, GPT-3.5 (Yu et al. \cite{HoneyGAN-YF}).

\begin{table}[h!]
\centering
\footnotesize
\caption{ \textit{Average success rate of \attackname\ for $1^{st}$, $3^{rd}$, $5^{th}$, and $10^{th}$ login attempts, with Honey-Chunk \cite{HoneyGAN-YF} HGT and 20 sweetwords per user account.}}
\label{tab:gpt-discussions-passgan}
\begin{tabular}{|c|cccc|}
\hline
\multirow{2}{*}{\textbf{Dataset}} & \multicolumn{4}{c|}{\textbf{Honey-Chunk (GPT-3.5)}}                                                     \\ \cline{2-5} 
 &
  \multicolumn{1}{c|}{\textbf{$1^{st}$ attempt}} &
  \multicolumn{1}{c|}{\textbf{$3^{rd}$ attempt}} &
  \multicolumn{1}{c|}{\textbf{$5^{th}$ attempt}} &
  \textbf{$10^{th}$ attempt} \\ \hline
\textbf{Adultfriendfinder} &
  \multicolumn{1}{c|}{8.89\%} &
  \multicolumn{1}{c|}{29.09\%} &
  \multicolumn{1}{c|}{42.63\%} &
  64.65\% \\ \hline
\textbf{MySpace}                  & \multicolumn{1}{c|}{15.60\%} & \multicolumn{1}{c|}{30.05\%} & \multicolumn{1}{c|}{41.97\%} & 66.51\% \\ \hline
\textbf{phpBB}                    & \multicolumn{1}{c|}{17.30\%} & \multicolumn{1}{c|}{23.74\%} & \multicolumn{1}{c|}{30.18\%} & 44.47\% \\ \hline
\textbf{LinkedIn}                 & \multicolumn{1}{c|}{7.38\%}  & \multicolumn{1}{c|}{19.47\%} & \multicolumn{1}{c|}{30.53\%} & 52.46\% \\ \hline
\textbf{Dropbox}                  & \multicolumn{1}{c|}{19.44\%} & \multicolumn{1}{c|}{39.48\%} & \multicolumn{1}{c|}{54.91\%} & 78.76\% \\ \hline
\textbf{RockYou-small}                  & \multicolumn{1}{c|}{16.15\%} & \multicolumn{1}{c|}{29.61\%} & \multicolumn{1}{c|}{39.96\%} & 64.39\% \\ \hline
\textbf{Dubsmash}                 & \multicolumn{1}{c|}{19.60\%} & \multicolumn{1}{c|}{45.60\%} & \multicolumn{1}{c|}{60.00\%} & 78.80\% \\ \hline
\textbf{Yahoo}                    & \multicolumn{1}{c|}{16.15\%} & \multicolumn{1}{c|}{29.61\%} & \multicolumn{1}{c|}{39.96\%} & 64.39\% \\ \hline
\textbf{Zynga}                    & \multicolumn{1}{c|}{13.82\%} & \multicolumn{1}{c|}{25.41\%} & \multicolumn{1}{c|}{35.16\%} & 53.46\% \\ \hline

\textbf{Average} &
  \multicolumn{1}{c|}{\textbf{14.92\%}} &
  \multicolumn{1}{c|}{\textbf{30.23\%}} &
  \multicolumn{1}{c|}{\textbf{41.70\%}} &
  \textbf{63.10\%} \\ \hline
\hline
\textbf{Random Guess} &
\multicolumn{1}{c|}{\textbf{5\%}} &
\multicolumn{1}{c|}{\textbf{15\%}} &
\multicolumn{1}{c|}{\textbf{25\%}} &
\textbf{50\%} \\ \hline
\end{tabular}%
\end{table}

The performance of \attackname\ when employing the GPT-3.5 for honeywords generation is depicted in Table~\ref{tab:gpt-discussions-passgan}. The results show that, in the self-trained attack settings with single login attempt the average success rate of password identification is 14.19\% significantly outperforming the 5\% chance agreement when associating 20 sweetwords with a user account. With increasing login attempts, we observed significant increase in success rate of attacker evading honeywords defense mechanism. Moreover, honeywords generated by the pre-trained LLM are similarly vulnerable to \attackname\ as those produced using a chaffing-with-a-password-model approach, where a FastText model is specifically trained for honeyword generation.

Our analysis of using GPT-3.5 for generating honeywords reveals an intriguing pattern (Table~\ref{tab:gpt-honeywords}). The GPT-3.5 model integrates more dictionary words into the honeywords, evident in instances where it flips characters and combines them with dictionary terms. For instance, from the password `sony1711,' it creates honeywords such as `SonyInnovate' and `SonyRevolution,' showcasing a blend of characters and dictionary words. This pattern enables the classifier to differentiate honeywords from passwords when using pre-trained models for honeyword generation.

\begin{table}[h!]
\centering
\footnotesize
\caption{\textit{Example of honeywords generated using Honey-Chunk (GPT-3.5 \cite{HoneyGAN-YF}).}}
\label{tab:gpt-honeywords}
\begin{tabular}{|ccc|}
\hline
\multicolumn{1}{|c|}{\textbf{sony1711 (Password)}} & \multicolumn{1}{c|}{SonyProgress} & FutureSony1711 \\ \hline
\multicolumn{1}{|c|}{Sony1711Connect} & \multicolumn{1}{c|}{Sony1711Network} & Sony1711Systems \\ \hline
\multicolumn{1}{|c|}{Sony1711Tech} & \multicolumn{1}{c|}{Sony1711Evolve} & Sony1711Explore \\ \hline
\end{tabular}%
\end{table}

\begin{tbox}[h!]
\begin{tcolorbox}[colframe=black!50!black, colbacktitle=black!40!white, 
coltitle=black, top=5pt, bottom=5pt, width=\columnwidth]
    In summary, our study shows that all the HGTs—including heuristics, representation learning, and LLM-based techniques—are vulnerable. The honeywords generated by these approaches can be effectively differentiated using our proposed attack, \attackname, employing a CNN model.
\end{tcolorbox}
\end{tbox}

%% file: discussions.tex
\section{Discussions}
\label{sec:discussions}
\noindent
In this section, we discuss the performance of \attackname\ in exceptional cases, its limitations, and future directions.

\smallskip 
\noindent
\textbf{Impact of Number of Sweetwords Per Account.} 
Juels and Rivest \cite{JA13} proposed that incorporating more ``sweetwords'' into user accounts could improve the security of systems that use password-based authentication. Our study compared the effectiveness of chaffing-with-a-password-model, and the chaffing-with-a-hybrid-model, when more than 20 sweetwords are associated with the user account. It was observed that the performance of our \attackname, decreased or remained similar to random guessing when using chaffing-with-a-password-model, but was significantly higher for chaffing-with-a-hybrid-model. 

\begin{table}[h!]
\centering
\footnotesize
\caption{\textit{Average success rate of \attackname\ in self-trained threat model in $1^{st}$ login attempt with chaffing-with-a-password-model and
chaffing-with-a-hybrid-model HGTs and 30 sweetwords per user account.}}

\label{tab:discussion-different-k}
\begin{tabular}{|c|cc|}
\hline
\multirow{2}{*}{\textbf{Dataset}} & \multicolumn{2}{c|}{\textbf{k = 30}}                                                                                                                                                       \\ \cline{2-3} 
                                  & \multicolumn{1}{c|}{\textbf{\begin{tabular}[c]{@{}c@{}}chaffing-with-a-\\ Password-Model\end{tabular}}} & \textbf{\begin{tabular}[c]{@{}c@{}}chaffing-with-a-\\ Hybrid-Model\end{tabular}} \\ \hline
\textbf{Adultfriendfinder}        & \multicolumn{1}{c|}{2.80\%}                                                                             & 29.60\%                                                                          \\ \hline
\textbf{MySpace}                  & \multicolumn{1}{c|}{2.40\%}                                                                             & 24.60\%                                                                          \\ \hline
\textbf{phpBB}                    & \multicolumn{1}{c|}{3.80\%}                                                                             & 39.20\%                                                                          \\ \hline
\textbf{LinkedIn}                 & \multicolumn{1}{c|}{3.00\%}                                                                             & 21.00\%                                                                          \\ \hline
\textbf{Dropbox}                  & \multicolumn{1}{c|}{2.80\%}                                                                             & 25.40\%                                                                          \\ \hline
\textbf{RockYou-small}                  & \multicolumn{1}{c|}{3.60\%}                                                                             & 29.80\%                                                                          \\ \hline
\textbf{Dubsmash}                 & \multicolumn{1}{c|}{2.80\%}                                                                             & 35.80\%                                                                          \\ \hline
\textbf{Yahoo}                    & \multicolumn{1}{c|}{2.00\%}                                                                             & 18.00\%                                                                          \\ \hline
\textbf{Zynga}                    & \multicolumn{1}{c|}{3.80\%}                                                                             & 29.20\%                                                                          \\ \hline
\textbf{Average}                  & \multicolumn{1}{c|}{\textbf{3.00\%}}                                                                    & \textbf{28.07\%}                                                                 \\ \hline
\end{tabular}%
\end{table}

Table~\ref{tab:discussion-different-k} shows the performance of \attackname\ when assigning 30 sweetwords to a user account. When generating honeywords using the chaffing-with-a-password-model HGT, the average success rate of \attackname\ is significantly lower as compared to random guessing. However, utilizing the chaffing-with-a-hybrid-model approach for honeyword generation elevates \attackname's success rate to 28.07\%, significantly surpassing random guessing (3.33\%). These results indicate that employing the chaffing-with-a-password-model method to generate honeywords can enhance the security of systems reliant on password-based authentication.

\smallskip
\noindent
\textbf{Impact of Numeric Passwords.} Some online passwords consist solely of numbers, representing the user's date of birth (DOB), phone number, area code, or other personal information. We found that generating honeywords for these numeric passwords results in sequences composed only of digits, posing a greater challenge for the CNN model to differentiate between passwords and sweetwords. Compared to numeric passwords, these honeywords could vary in length, either shorter or longer. Under these circumstances, we observed a significant decline in the CNN model's performance. Using the chaffing-with-a-password model, the classifier achieved an average success rate of 0.8\%, and 1.2\%. with 20 sweetwords linked to a user account.

\smallskip
\noindent
\textbf{Mitigation Strategies.} To counteract \attackname, organizations should enforce the generation of randomized passwords using password managers and incorporate two-factor authentication (2FA) as mandatory practice (if not already). These strategies effectively complicate the ability of attackers to perform ODA. Specifically as shown in \cite{MSSJNSNN-14}, the integration of 2FA, where a salted hash of the password is stored on the server and the corresponding salt is kept on the 2FA device, renders ODA impractical. This setup ensures that the critical components required for authentication are separated and independently secured, significantly enhancing the overall security of the system.

\smallskip
\noindent
\textbf{Limitations.} In this study, we focus on passwords and honeywords tied to user accounts, without considering PII. However, future research could beneficially explore the impact of incorporating PII in identifying user passwords, and in the process of honeywords generation in a way that privacy of the user is preserved. Additionally, the use of passwords created by password managers (e.g. 1Password) or those generated by built-in browser managers (e.g., Chrome) can also be examined. These methods produce random passwords devoid of PII, rendering them more secure against targeted guessing attacks \cite{DW18}, wherein attackers exploit a user’s PII to deduce their password.

\smallskip
\noindent
\textbf{Adapting \attackname\ to Break Honey Encryption (HE).} The application of \attackname\ on honeywords suggests potential adaptability to HE systems. HE systems, like honeywords, use decoy outputs to confuse attackers. This similarity presents a promising opportunity to apply our developed attack methodologies to HE. Employing DL models to identify statistical patterns and exploit differences in entropy between genuine and decoy outputs could significantly advance cybersecurity practices. Further research should focus on exploring these vulnerabilities in HE, guided by foundational studies that highlight the effectiveness and challenges of decoy strategies \cite{GMBB-16-HE, FDDCJ-24-HE}.

\smallskip
\noindent
\textbf{Future Work.} Future studies should evaluate \attackname's performance with alternative DNNs, including long-short term memory (LSTM), stacked denoising autoencoders (SDAE), and other notable CNN architectures. Additionally, exploring the effect of our attack with honeywords generated by other language models like Bidirectional Encoder Representations from Transformers (BERT) \cite{BERT}, Global Vectors for Word Representation (GLoVe) \cite{glove}, or GPT \cite{chatgpt-paper} fine-tuned for honeyword generation not the the pre-traiend models considered in existing approaches for honeywords generation would be compelling. Additionally, the effectiveness of using a Bernoulli process for selecting honeywords \cite{WKCRMK-24} should be assessed. 

%% file: related-work.tex
\section{Related Work}
\label{lab:related-work}
\noindent
Jules and Rivest \cite{JA13} introduced the concept of honeywords to enhance the security of hashed passwords and suggested multiple techniques for their generation, including tweaking algorithms and model-based approaches \cite{HB10-kamouflage, WM09SM}. They advocated for the use of 19 honeywords per account. Subsequently, Wang et al. \cite{DW18} applied these strategies to large datasets but found them inadequate, with attack success rates ranging from 29.29\% to 67.9\% across different scenarios. They developed an alternate method, utilizing PII to generate honeywords, reducing the success rate to 18.2\% \cite{DWZY22}.

Wang and Reiter \cite{KCW21Amnesia} proposed Amnesia, a technique that uses probabilistic marking to detect breaches without persistent state. Conversely, Dionysiou and Elias \cite{DA22Lethe} noted limitations with Amnesia and proposed Lethe as an alternative, using honeywords to detect data breaches without needing a trusted component to distinguish passwords from user input credentials. Erguler et al. \cite{EI16} developed `honeyindex,' which employs other users' passwords as honeywords to minimize storage needs, though it faced deployment challenges \cite{DWZY22}. Chakraborty et al. \cite{NC18} proposed Paired Distance Protocol (PDP), which enhances security by requiring users to remember a random string alongside their password, though it's susceptible to Multiple System Intersection attacks \cite{AC19}. Akshima et al. \cite{AC19} studied advanced honeyword generation techniques (e.g., evolving-password-model, user-profile-model, append-secret-model) which prove resistant to certain attacks but vulnerable to others \cite{DAWZY22}. Guo et al. \cite{GY21} introduced Superwords, a method that disconnects usernames from passwords, significantly reducing the probability of successful attacks \cite{JA13}.

%% file: conclusion.tex
\section{Conclusion}
\label{sec:conclusion}
\noindent
In conclusion, our research introduces \attackname, a novel DL-based attack that utilized CNN architecture to identify real passwords from a set of sweetwords, and adapt  $\epsilon$-flatness metric for DL-based attacks that utilized model-predicted probabilities to rank the sweetwords for password login attempts. Moreover, we also proposed three threat models that represent real-world attack scenarios. These threat models -- the same-service, cross-service, and self-trained threat models -- contribute to various attack vectors that underscores the relevance of our findings in practical security settings. Our results show that \attackname\ can effectively compromise the security of HGTs proposed by Dionysiou et al. \cite{DA21} and Yu et al. \cite{HoneyGAN-YF} without requiring any PII. This robust performance of \attackname\ highlights that existing HGTs are vulnerable and underscores the need for more secure approaches. Furthermore, our comparative analysis shows that honeywords generated using general-purpose language models (e.g., GPT-3.5 \cite{HoneyGAN-YF}) are more susceptible to \attackname\ as compared to those generated using specialized model (chaffing-with-a-password-model \cite{DA21}). Consequently, we advocate for the use of dedicated models specifically trained for honeyword generation, that can help enhance their resilience against sophisticated attacks.

%% file: appendix.tex
\section{Results of \attackname\ on individual datasets}
\label{lab:exp-3-5-10-attempts}
In this section, we provide supplementary material for experiments conducted for each of the threat models. The results presents the $\epsilon$-flatness (success-rate) of \attackname\ when attacker is allowed more than 1 attempt to log in to the system. Table~\ref{tab:exp-a-1-3-5-10-attempts}, \ref{tab:exp-b-1-3-5-10-attempts}, and \ref{tab:exp-c-1-3-5-10-attempts} shows the average success rate of \attackname\ for password identification when 20 sweetwords are associated with the user account when attacker make 1, 3, 5, or 10 login attempts.

\begin{table*}[h!]
\centering
\small
\caption{\textit{Average success rate of \attackname\ in \textbf{Experiment A} in $1^{st}$, $3^{rd}$, $5^{th}$, and $10^{th}$ login attempt, under \textbf{same-service threat model} with different HGTs and 20 sweetwords per user account.}}
\label{tab:exp-a-1-3-5-10-attempts}
\resizebox{\textwidth}{!}{%
\begin{tabular}{|c|cccc|cccc|cccc|}
\hline
 &
  \multicolumn{4}{c|}{\textbf{chaffing-by-tweaking}} &
  \multicolumn{4}{c|}{\textbf{chaffing-with-a-password-model}} &
  \multicolumn{4}{c|}{\textbf{chaffing-with-a-hybrid-model}} \\ \hline
\textbf{Dataset} &
  \multicolumn{1}{c|}{$1^{st}$} &
  \multicolumn{1}{c|}{$3^{rd}$} &
  \multicolumn{1}{c|}{$5^{th}$} &
  $10^{th}$ &
  \multicolumn{1}{c|}{$1^{st}$} &
  \multicolumn{1}{c|}{$3^{rd}$} &
  \multicolumn{1}{c|}{$5^{th}$} &
  $10^{th}$ &
  \multicolumn{1}{c|}{$1^{st}$} &
  \multicolumn{1}{c|}{$3^{rd}$} &
  \multicolumn{1}{c|}{$5^{th}$} &
  $10^{th}$ \\ \hline
\textbf{Adultfriendfinder} &
  \multicolumn{1}{c|}{54.80\%} &
  \multicolumn{1}{c|}{83.60\%} &
  \multicolumn{1}{c|}{95.00\%} &
  100.00\% &
  \multicolumn{1}{c|}{17.20\%} &
  \multicolumn{1}{c|}{38.40\%} &
  \multicolumn{1}{c|}{55.20\%} &
  80.20\% &
  \multicolumn{1}{c|}{57.20\%} &
  \multicolumn{1}{c|}{84.80\%} &
  \multicolumn{1}{c|}{95.60\%} &
  100.00\% \\ \hline
\textbf{MySpace} &
  \multicolumn{1}{c|}{63.00\%} &
  \multicolumn{1}{c|}{86.60\%} &
  \multicolumn{1}{c|}{93.60\%} &
  98.80\% &
  \multicolumn{1}{c|}{11.00\%} &
  \multicolumn{1}{c|}{23.00\%} &
  \multicolumn{1}{c|}{35.00\%} &
  69.40\% &
  \multicolumn{1}{c|}{55.40\%} &
  \multicolumn{1}{c|}{82.40\%} &
  \multicolumn{1}{c|}{90.00\%} &
  98.20\% \\ \hline
\textbf{phpBB} &
  \multicolumn{1}{c|}{63.00\%} &
  \multicolumn{1}{c|}{83.60\%} &
  \multicolumn{1}{c|}{90.40\%} &
  97.60\% &
  \multicolumn{1}{c|}{15.60\%} &
  \multicolumn{1}{c|}{43.00\%} &
  \multicolumn{1}{c|}{61.20\%} &
  88.00\% &
  \multicolumn{1}{c|}{63.80\%} &
  \multicolumn{1}{c|}{89.40\%} &
  \multicolumn{1}{c|}{96.80\%} &
  99.80\% \\ \hline
\textbf{LinkedIn} &
  \multicolumn{1}{c|}{43.80\%} &
  \multicolumn{1}{c|}{74.00\%} &
  \multicolumn{1}{c|}{90.60\%} &
  99.40\% &
  \multicolumn{1}{c|}{14.60\%} &
  \multicolumn{1}{c|}{37.60\%} &
  \multicolumn{1}{c|}{53.40\%} &
  82.40\% &
  \multicolumn{1}{c|}{28.80\%} &
  \multicolumn{1}{c|}{69.20\%} &
  \multicolumn{1}{c|}{87.80\%} &
  98.20\% \\ \hline
\textbf{Dropbox} &
  \multicolumn{1}{c|}{48.20\%} &
  \multicolumn{1}{c|}{77.40\%} &
  \multicolumn{1}{c|}{88.80\%} &
  97.20\% &
  \multicolumn{1}{c|}{16.20\%} &
  \multicolumn{1}{c|}{38.60\%} &
  \multicolumn{1}{c|}{51.80\%} &
  80.80\% &
  \multicolumn{1}{c|}{49.20\%} &
  \multicolumn{1}{c|}{76.00\%} &
  \multicolumn{1}{c|}{90.40\%} &
  99.20\% \\ \hline
\textbf{RockYou} &
  \multicolumn{1}{c|}{59.80\%} &
  \multicolumn{1}{c|}{84.20\%} &
  \multicolumn{1}{c|}{92.20\%} &
  98.60\% &
  \multicolumn{1}{c|}{4.80\%} &
  \multicolumn{1}{c|}{20.60\%} &
  \multicolumn{1}{c|}{35.20\%} &
  68.60\% &
  \multicolumn{1}{c|}{56.40\%} &
  \multicolumn{1}{c|}{80.80\%} &
  \multicolumn{1}{c|}{90.00\%} &
  98.40\% \\ \hline
\textbf{Dubsmash} &
  \multicolumn{1}{c|}{63.20\%} &
  \multicolumn{1}{c|}{84.00\%} &
  \multicolumn{1}{c|}{94.20\%} &
  99.60\% &
  \multicolumn{1}{c|}{13.00\%} &
  \multicolumn{1}{c|}{33.40\%} &
  \multicolumn{1}{c|}{51.00\%} &
  81.80\% &
  \multicolumn{1}{c|}{59.60\%} &
  \multicolumn{1}{c|}{83.40\%} &
  \multicolumn{1}{c|}{92.80\%} &
  98.80\% \\ \hline
\textbf{Yahoo} &
  \multicolumn{1}{c|}{30.40\%} &
  \multicolumn{1}{c|}{57.60\%} &
  \multicolumn{1}{c|}{75.40\%} &
  93.40\% &
  \multicolumn{1}{c|}{24.20\%} &
  \multicolumn{1}{c|}{48.80\%} &
  \multicolumn{1}{c|}{65.00\%} &
  88.60\% &
  \multicolumn{1}{c|}{43.20\%} &
  \multicolumn{1}{c|}{74.60\%} &
  \multicolumn{1}{c|}{87.60\%} &
  99.00\% \\ \hline
\textbf{Zynga} &
  \multicolumn{1}{c|}{48.80\%} &
  \multicolumn{1}{c|}{82.00\%} &
  \multicolumn{1}{c|}{94.80\%} &
  99.00\% &
  \multicolumn{1}{c|}{16.80\%} &
  \multicolumn{1}{c|}{38.40\%} &
  \multicolumn{1}{c|}{56.40\%} &
  83.60\% &
  \multicolumn{1}{c|}{51.00\%} &
  \multicolumn{1}{c|}{80.60\%} &
  \multicolumn{1}{c|}{92.00\%} &
  98.80\% \\ \hline
\textbf{Average} &
  \multicolumn{1}{c|}{\textbf{52.78\%}} &
  \multicolumn{1}{c|}{\textbf{79.22\%}} &
  \multicolumn{1}{c|}{\textbf{90.56\%}} &
  \textbf{98.18\%} &
  \multicolumn{1}{c|}{\textbf{14.82\%}} &
  \multicolumn{1}{c|}{\textbf{35.76\%}} &
  \multicolumn{1}{c|}{\textbf{51.58\%}} &
  \textbf{80.38\%} &
  \multicolumn{1}{c|}{\textbf{51.62\%}} &
  \multicolumn{1}{c|}{\textbf{80.13\%}} &
  \multicolumn{1}{c|}{\textbf{91.44\%}} &
  \textbf{98.93\%} \\ \hline
\end{tabular}%
}
\end{table*}

\begin{table*}[h!]
\centering
\small
\caption{\textit{Average success rate of \attackname\ in \textbf{Experiment B} in $1^{st}$, $3^{rd}$, $5^{th}$, and $10^{th}$ login attempt, under \textbf{cross-service threat model} with different HGTs and 20 sweetwords per user account.}}
\label{tab:exp-b-1-3-5-10-attempts}
\resizebox{\textwidth}{!}{%
\begin{tabular}{|c|cccc|cccc|cccc|}
\hline
 &
  \multicolumn{4}{c|}{\textbf{chaffing-by-tweaking}} &
  \multicolumn{4}{c|}{\textbf{chaffing-with-a-password-model}} &
  \multicolumn{4}{c|}{\textbf{chaffing-with-a-hybrid-model}} \\ \hline
\textbf{Dataset} &
  \multicolumn{1}{c|}{$1^{st}$} &
  \multicolumn{1}{c|}{$3^{rd}$} &
  \multicolumn{1}{c|}{$5^{th}$} &
  $10^{th}$ &
  \multicolumn{1}{c|}{\textbf{$1^{st}$}} &
  \multicolumn{1}{c|}{$3^{rd}$} &
  \multicolumn{1}{c|}{$5^{th}$} &
  $10^{th}$ &
  \multicolumn{1}{c|}{$1^{st}$} &
  \multicolumn{1}{c|}{$3^{rd}$} &
  \multicolumn{1}{c|}{$5^{th}$} &
  $10^{th}$ \\ \hline
\textbf{Adultfriendfinder} &
  \multicolumn{1}{c|}{49.80\%} &
  \multicolumn{1}{c|}{76.60\%} &
  \multicolumn{1}{c|}{90.80\%} &
  99.40\% &
  \multicolumn{1}{c|}{6.80\%} &
  \multicolumn{1}{c|}{24.40\%} &
  \multicolumn{1}{c|}{40.80\%} &
  75.00\% &
  \multicolumn{1}{c|}{48.20\%} &
  \multicolumn{1}{c|}{78.40\%} &
  \multicolumn{1}{c|}{89.80\%} &
  99.20\% \\ \hline
\textbf{MySpace} &
  \multicolumn{1}{c|}{52.60\%} &
  \multicolumn{1}{c|}{78.20\%} &
  \multicolumn{1}{c|}{89.40\%} &
  97.40\% &
  \multicolumn{1}{c|}{9.60\%} &
  \multicolumn{1}{c|}{26.80\%} &
  \multicolumn{1}{c|}{39.60\%} &
  69.60\% &
  \multicolumn{1}{c|}{47.40\%} &
  \multicolumn{1}{c|}{76.80\%} &
  \multicolumn{1}{c|}{87.80\%} &
  97.60\% \\ \hline
\textbf{phpBB} &
  \multicolumn{1}{c|}{61.20\%} &
  \multicolumn{1}{c|}{80.60\%} &
  \multicolumn{1}{c|}{90.40\%} &
  97.80\% &
  \multicolumn{1}{c|}{6.00\%} &
  \multicolumn{1}{c|}{24.00\%} &
  \multicolumn{1}{c|}{40.20\%} &
  74.80\% &
  \multicolumn{1}{c|}{54.80\%} &
  \multicolumn{1}{c|}{80.20\%} &
  \multicolumn{1}{c|}{90.60\%} &
  99.20\% \\ \hline
\textbf{LinkedIn} &
  \multicolumn{1}{c|}{43.20\%} &
  \multicolumn{1}{c|}{72.80\%} &
  \multicolumn{1}{c|}{88.80\%} &
  97.60\% &
  \multicolumn{1}{c|}{6.20\%} &
  \multicolumn{1}{c|}{24.40\%} &
  \multicolumn{1}{c|}{41.80\%} &
  68.80\% &
  \multicolumn{1}{c|}{44.20\%} &
  \multicolumn{1}{c|}{68.80\%} &
  \multicolumn{1}{c|}{82.00\%} &
  95.80\% \\ \hline
\textbf{Dropbox} &
  \multicolumn{1}{c|}{49.60\%} &
  \multicolumn{1}{c|}{77.00\%} &
  \multicolumn{1}{c|}{88.60\%} &
  95.80\% &
  \multicolumn{1}{c|}{10.40\%} &
  \multicolumn{1}{c|}{28.40\%} &
  \multicolumn{1}{c|}{45.00\%} &
  74.20\% &
  \multicolumn{1}{c|}{45.80\%} &
  \multicolumn{1}{c|}{73.80\%} &
  \multicolumn{1}{c|}{88.80\%} &
  98.40\% \\ \hline
\textbf{RockYou} &
  \multicolumn{1}{c|}{56.20\%} &
  \multicolumn{1}{c|}{77.80\%} &
  \multicolumn{1}{c|}{88.20\%} &
  97.40\% &
  \multicolumn{1}{c|}{5.00\%} &
  \multicolumn{1}{c|}{18.80\%} &
  \multicolumn{1}{c|}{30.40\%} &
  62.60\% &
  \multicolumn{1}{c|}{48.00\%} &
  \multicolumn{1}{c|}{73.00\%} &
  \multicolumn{1}{c|}{83.20\%} &
  96.40\% \\ \hline
\textbf{Dubsmash} &
  \multicolumn{1}{c|}{63.20\%} &
  \multicolumn{1}{c|}{84.00\%} &
  \multicolumn{1}{c|}{94.20\%} &
  99.60\% &
  \multicolumn{1}{c|}{13.00\%} &
  \multicolumn{1}{c|}{33.40\%} &
  \multicolumn{1}{c|}{51.00\%} &
  81.80\% &
  \multicolumn{1}{c|}{59.60\%} &
  \multicolumn{1}{c|}{83.40\%} &
  \multicolumn{1}{c|}{92.80\%} &
  98.80\% \\ \hline
\textbf{Yahoo} &
  \multicolumn{1}{c|}{40.40\%} &
  \multicolumn{1}{c|}{68.40\%} &
  \multicolumn{1}{c|}{79.80\%} &
  94.60\% &
  \multicolumn{1}{c|}{12.20\%} &
  \multicolumn{1}{c|}{30.80\%} &
  \multicolumn{1}{c|}{46.80\%} &
  73.80\% &
  \multicolumn{1}{c|}{37.40\%} &
  \multicolumn{1}{c|}{67.20\%} &
  \multicolumn{1}{c|}{83.20\%} &
  95.40\% \\ \hline
\textbf{Zynga} &
  \multicolumn{1}{c|}{50.40\%} &
  \multicolumn{1}{c|}{81.40\%} &
  \multicolumn{1}{c|}{93.40\%} &
  99.40\% &
  \multicolumn{1}{c|}{4.80\%} &
  \multicolumn{1}{c|}{22.40\%} &
  \multicolumn{1}{c|}{40.40\%} &
  75.20\% &
  \multicolumn{1}{c|}{47.80\%} &
  \multicolumn{1}{c|}{78.20\%} &
  \multicolumn{1}{c|}{89.40\%} &
  99.00\% \\ \hline
\textbf{Average} &
  \multicolumn{1}{c|}{\textbf{51.84\%}} &
  \multicolumn{1}{c|}{\textbf{77.42\%}} &
  \multicolumn{1}{c|}{\textbf{89.29\%}} &
  \textbf{97.67\%} &
  \multicolumn{1}{c|}{\textbf{8.22\%}} &
  \multicolumn{1}{c|}{\textbf{25.93\%}} &
  \multicolumn{1}{c|}{\textbf{41.78\%}} &
  \textbf{72.87\%} &
  \multicolumn{1}{c|}{\textbf{48.13\%}} &
  \multicolumn{1}{c|}{\textbf{75.53\%}} &
  \multicolumn{1}{c|}{\textbf{87.51\%}} &
  \textbf{97.76\%} \\ \hline
\end{tabular}%
}
\end{table*}

\begin{table*}[h!]
\centering
\small
\caption{\textit{Average success rate of \attackname\ in \textbf{Experiment C} in $1^{st}$, $3^{rd}$, $5^{th}$, and $10^{th}$ login attempt, under \textbf{self-trained threat model} with different HGTs and 20 sweetwords per user account.}}
\label{tab:exp-c-1-3-5-10-attempts}
\resizebox{\textwidth}{!}{%
\begin{tabular}{|c|cccc|cccc|cccc|}
\hline
\textbf{} &
  \multicolumn{4}{c|}{\textbf{chaffing-by-tweaking}} &
  \multicolumn{4}{c|}{\textbf{chaffing-with-a-password-model}} &
  \multicolumn{4}{c|}{\textbf{chaffing-with-a-hybrid-model}} \\ \hline
\textbf{Dataset} &
  \multicolumn{1}{c|}{$1^{st}$} &
  \multicolumn{1}{c|}{$3^{rd}$} &
  \multicolumn{1}{c|}{$5^{th}$} &
  $10^{th}$ &
  \multicolumn{1}{c|}{$1^{st}$} &
  \multicolumn{1}{c|}{$3^{rd}$} &
  \multicolumn{1}{c|}{$5^{th}$} &
  $10^{th}$ &
  \multicolumn{1}{c|}{$1^{st}$} &
  \multicolumn{1}{c|}{$3^{rd}$} &
  \multicolumn{1}{c|}{$5^{th}$} &
  $10^{th}$ \\ \hline
\textbf{Adultfriendfinder} &
  \multicolumn{1}{c|}{48.40\%} &
  \multicolumn{1}{c|}{76.40\%} &
  \multicolumn{1}{c|}{89.80\%} &
  99.60\% &
  \multicolumn{1}{c|}{4.00\%} &
  \multicolumn{1}{c|}{13.00\%} &
  \multicolumn{1}{c|}{21.00\%} &
  50.00\% &
  \multicolumn{1}{c|}{32.20\%} &
  \multicolumn{1}{c|}{63.80\%} &
  \multicolumn{1}{c|}{77.40\%} &
  94.80\% \\ \hline
\textbf{MySpace} &
  \multicolumn{1}{c|}{43.00\%} &
  \multicolumn{1}{c|}{70.60\%} &
  \multicolumn{1}{c|}{82.60\%} &
  95.80\% &
  \multicolumn{1}{c|}{3.40\%} &
  \multicolumn{1}{c|}{10.60\%} &
  \multicolumn{1}{c|}{19.20\%} &
  42.40\% &
  \multicolumn{1}{c|}{29.80\%} &
  \multicolumn{1}{c|}{53.40\%} &
  \multicolumn{1}{c|}{64.00\%} &
  83.60\% \\ \hline
\textbf{phpBB} &
  \multicolumn{1}{c|}{56.20\%} &
  \multicolumn{1}{c|}{74.80\%} &
  \multicolumn{1}{c|}{83.40\%} &
  94.00\% &
  \multicolumn{1}{c|}{5.20\%} &
  \multicolumn{1}{c|}{14.20\%} &
  \multicolumn{1}{c|}{25.60\%} &
  59.80\% &
  \multicolumn{1}{c|}{45.40\%} &
  \multicolumn{1}{c|}{67.60\%} &
  \multicolumn{1}{c|}{78.00\%} &
  90.60\% \\ \hline
\textbf{LinkedIn} &
  \multicolumn{1}{c|}{39.60\%} &
  \multicolumn{1}{c|}{68.40\%} &
  \multicolumn{1}{c|}{81.40\%} &
  93.80\% &
  \multicolumn{1}{c|}{4.80\%} &
  \multicolumn{1}{c|}{13.20\%} &
  \multicolumn{1}{c|}{21.00\%} &
  47.80\% &
  \multicolumn{1}{c|}{29.20\%} &
  \multicolumn{1}{c|}{51.20\%} &
  \multicolumn{1}{c|}{66.00\%} &
  85.60\% \\ \hline
\textbf{Dropbox} &
  \multicolumn{1}{c|}{41.40\%} &
  \multicolumn{1}{c|}{67.00\%} &
  \multicolumn{1}{c|}{78.80\%} &
  93.40\% &
  \multicolumn{1}{c|}{5.00\%} &
  \multicolumn{1}{c|}{14.00\%} &
  \multicolumn{1}{c|}{22.00\%} &
  53.00\% &
  \multicolumn{1}{c|}{32.00\%} &
  \multicolumn{1}{c|}{57.20\%} &
  \multicolumn{1}{c|}{68.60\%} &
  87.00\% \\ \hline
\textbf{RockYou} &
  \multicolumn{1}{c|}{49.80\%} &
  \multicolumn{1}{c|}{74.20\%} &
  \multicolumn{1}{c|}{84.40\%} &
  94.80\% &
  \multicolumn{1}{c|}{5.40\%} &
  \multicolumn{1}{c|}{14.00\%} &
  \multicolumn{1}{c|}{22.20\%} &
  46.60\% &
  \multicolumn{1}{c|}{33.00\%} &
  \multicolumn{1}{c|}{58.60\%} &
  \multicolumn{1}{c|}{72.00\%} &
  89.40\% \\ \hline
\textbf{Dubsmash} &
  \multicolumn{1}{c|}{53.20\%} &
  \multicolumn{1}{c|}{77.00\%} &
  \multicolumn{1}{c|}{87.00\%} &
  98.20\% &
  \multicolumn{1}{c|}{6.00\%} &
  \multicolumn{1}{c|}{16.80\%} &
  \multicolumn{1}{c|}{28.40\%} &
  59.80\% &
  \multicolumn{1}{c|}{42.00\%} &
  \multicolumn{1}{c|}{64.00\%} &
  \multicolumn{1}{c|}{73.60\%} &
  87.40\% \\ \hline
\textbf{Yahoo} &
  \multicolumn{1}{c|}{32.60\%} &
  \multicolumn{1}{c|}{54.00\%} &
  \multicolumn{1}{c|}{66.60\%} &
  86.80\% &
  \multicolumn{1}{c|}{3.20\%} &
  \multicolumn{1}{c|}{9.40\%} &
  \multicolumn{1}{c|}{16.60\%} &
  46.80\% &
  \multicolumn{1}{c|}{23.40\%} &
  \multicolumn{1}{c|}{41.00\%} &
  \multicolumn{1}{c|}{52.40\%} &
  75.80\% \\ \hline
\textbf{Zynga} &
  \multicolumn{1}{c|}{46.20\%} &
  \multicolumn{1}{c|}{73.00\%} &
  \multicolumn{1}{c|}{85.60\%} &
  98.20\% &
  \multicolumn{1}{c|}{5.20\%} &
  \multicolumn{1}{c|}{15.20\%} &
  \multicolumn{1}{c|}{26.80\%} &
  57.40\% &
  \multicolumn{1}{c|}{36.00\%} &
  \multicolumn{1}{c|}{58.40\%} &
  \multicolumn{1}{c|}{72.60\%} &
  88.40\% \\ \hline
\textbf{PassGAN} &
  \multicolumn{1}{c|}{68.60\%} &
  \multicolumn{1}{c|}{90.20\%} &
  \multicolumn{1}{c|}{96.60\%} &
  99.40\% &
  \multicolumn{1}{c|}{18.80\%} &
  \multicolumn{1}{c|}{41.60\%} &
  \multicolumn{1}{c|}{55.00\%} &
  83.60\% &
  \multicolumn{1}{c|}{66.00\%} &
  \multicolumn{1}{c|}{89.80\%} &
  \multicolumn{1}{c|}{96.80\%} &
  99.00\% \\ \hline
\textbf{Average} &
  \multicolumn{1}{c|}{\textbf{68.60\%}} &
  \multicolumn{1}{c|}{\textbf{90.20\%}} &
  \multicolumn{1}{c|}{\textbf{96.60\%}} &
  \textbf{99.40\%} &
  \multicolumn{1}{c|}{\textbf{18.80\%}} &
  \multicolumn{1}{c|}{\textbf{41.60\%}} &
  \multicolumn{1}{c|}{\textbf{55.00\%}} &
  \textbf{83.60\%} &
  \multicolumn{1}{c|}{\textbf{66.00\%}} &
  \multicolumn{1}{c|}{\textbf{89.80\%}} &
  \multicolumn{1}{c|}{\textbf{96.80\%}} &
  \textbf{99.00\%} \\ \hline
\end{tabular}%
}
\end{table*}